\documentclass[sigconf,nonacm,authorversion]{acmart}
\AtBeginDocument{%
  }

\setcopyright{acmlicensed}
\copyrightyear{2018}
\acmYear{2018}
\acmDOI{XXXXXXX.XXXXXXX}
\acmConference[Conference acronym 'XX]{Make sure to enter the correct
  conference title from your rights confirmation email}{June 03--05,
  2018}{Woodstock, NY}
\acmISBN{978-1-4503-XXXX-X/2018/06}

\usepackage{balance}
\usepackage{comment}
\usepackage{multirow}
\usepackage{booktabs}

\usepackage{amsmath,amssymb,amsfonts}
\usepackage{algorithmic}
\usepackage{graphicx}
\usepackage{float}
\usepackage{textcomp}
\usepackage{url}
\usepackage{hyperref}
\usepackage{tcolorbox}
\usepackage{xcolor}
\usepackage{caption}
\usepackage{stfloats}
\usepackage{url}
\begin{document}

\title{Do We Talk to Robots Like Therapists, and Do They Respond Accordingly? Language Alignment in AI Emotional Support}

\author{Sophie Chiang}
\orcid{ }
\affiliation{%
  \institution{Department of Computer Science and Technology,\\ University of Cambridge}
  \city{Cambridge}
  \country{UK}}
\email{ }

\author{Guy Laban}
\orcid{0000-0002-3796-1804}
\affiliation{%
  \institution{Department of Computer Science and Technology,\\ University of Cambridge}
  \city{Cambridge}
  \country{UK}}
\email{guy.laban@cl.cam.ac.uk}

\author{Hatice Gunes}
\orcid{ }
\affiliation{%
  \institution{Department of Computer Science and Technology,\\ University of Cambridge}
  \city{Cambridge}
  \country{UK}}
\email{ }


\begin{abstract}

As conversational agents increasingly engage in emotionally supportive dialogue, it is important to understand how closely their interactions resemble those in traditional therapy settings. This study investigates whether the concerns shared with a robot align with those shared in human-to-human (H2H) therapy sessions, and whether robot responses semantically mirror those of human therapists. We analysed two datasets: one of interactions between users and professional therapists (Hugging Face’s NLP Mental Health Conversations), and another involving supportive conversations with a social robot (QTrobot; LuxAI) powered by a large language model (LLM; GPT 3.5). Using sentence embeddings and K-means clustering, 
we assessed cross-agent thematic alignment by applying a distance-based cluster-fitting method that evaluates whether responses from one agent type map to clusters derived from the other, and validating it using euclidean distances. Results showed that 90.88\% of robot conversation disclosures could be mapped to clusters from the human therapy dataset, suggesting shared topical structure. For matched clusters, we compared the subjects as well as therapist and robot responses using Transformer, Word2Vec, and BERT embeddings, revealing strong semantic overlap in subjects' disclosures in both datasets, as well as in the responses given to similar human disclosure themes across agent types (robot vs. human therapist). These findings highlight both the parallels and boundaries of robot-led support conversations and their potential for augmenting mental health interventions.

  
\end{abstract}

\ccsdesc[500]{Human-centered computing~HCI theory, concepts and models}
\ccsdesc[500]{Human-centered computing~Natural language interfaces}
\ccsdesc[500]{Human-centered computing~Empirical studies in HCI}

\keywords{Human--Robot Interaction, Emotional Support, Semantic alignment, Self-disclosure, Large language models}

\maketitle

\section{Introduction}

As the role of conversational artificial agents (e.g., social robots) in providing emotional support is increasingly discussed \cite{RefWorks:404,Laban2022SocialTreatment}, questions arise regarding how well these systems replicate, augment, or diverge from human therapeutic dialogue. Thanks to recent advances in conversational AI, such as large language models (LLMs), these systems are now capable of holding thoughtful, sensitive conversations that go beyond basic question-and-answer exchanges and rule-based interactions \cite{Laban2024SharingFeel}. When embodied in physically present artificial agents like social robots, these systems go beyond basic communication by offering a sense of presence and rapport that can feel personal and emotionally engaging  \cite{Henschel2021,Laban2024SharingFeel}. Yet, despite the expanding capabilities of social robots \cite{Spitale2024PastWell-being}, and the growing interests in robot-led interactions in well-being interventions \cite{Laban2024SocialWell-Being,RefWorks:424}, there remains limited understanding of how closely these interactions align, both semantically and thematically, with those in human therapeutic contexts.

Prior work has shown that people tend to open up towards these robots \cite{Laban2023OpeningBehavior} disclosing gradually more towards them overtime \cite{Laban2024BuildingTime,laban_ced_2023,laban2025robot}, especially when they perceive the interaction as nonjudgmental or confidential \cite{Laban2024SharingFeel}. Moreover, people seem to talk about similar topics across interactions with humans, robots, and disembodied artificial agents in general everyday settings \cite{Chiang2025ComparingAgent}. However, it remains unclear whether the topics raised with robots in care and support settings align with those shared with human professionals. Examining whether the topics (i.e., themes of disclosure) people raise with a social robot are comparable to, or align with, those shared with a human therapist can help us better understand whether people seek similar forms of support from robots. Understanding this overlap, or divergence, can help determine the depth and relevance of robot-led support and how it might be perceived by human users according to their desired topics of discussion. 
Accordingly, we are asking-
\textbf{RQ1:} \textit{To what extent people's topics of disclosure to a social robot align with those to a human therapist?}

Beyond thematic alignment, it is also crucial to consider how participants articulate their disclosure across contexts. Even when similar topics are discussed, the way users express themselves 
can vary depending on the conversational setting and the perceived role of the agent, especially when the agent is a robot aiming to take a supportive role. Prior work suggests that people tend to adapt their verbal communication behaviour based on the perceived social attributes of their interlocutor or the situation (e.g., \cite{Gonzales2010LanguageGroups,Ireland2011LanguageStability, Niederhoffer2002LinguisticInteraction, Aafjes-vanDoorn2020LanguageAlliance}), a phenomenon captured by frameworks such as Communication Accommodation Theory (CAT) \cite{giles1991contexts} and Language Style Matching (LSM) \cite{Gonzales2010LanguageGroups,Gasiorek2021InteractionalPsychology,Niederhoffer2002LinguisticInteraction}. This raises the question of whether similar themes of disclosure are expressed in comparable ways during supportive interactions with a robotic agent versus a human therapist, or whether differences emerge in how people construct their narratives. By examining the semantic similarity of participants' disclosures across human-to-human (therapist) and human-to-robot interactions, we can evaluate whether people convey similar semantic content and narrative structures, or if differences emerge in how they frame their experiences. This analysis offers insight into whether robot-led support elicits comparable articulations and personal disclosures, and whether users communicate their needs in comparable ways across modalities. Accordingly, we are asking - \textbf{RQ2:} \textit{When people express similar themes of disclosure, to what extent are their disclosures semantically similar across interactions with human therapists and a robotic agent?}

While understanding how users articulate their disclosures is essential, it is equally important to examine how these disclosures are received and addressed by the agents themselves, highlighting the content that is communicated during emotionally supportive interactions, as well as how the supportive interlocutor (whether a human therapist or a robot) responds in ways that could feel attuned and supportive \cite{Gelso1985ThePsychotherapy}. If robot-led support is to be taken seriously as an intervention for emotional care, it is crucial to assess whether its responses reflect the kinds of empathic and semantically meaningful patterns found in human therapy \cite{Laban2025CriticalWellbeing}. To this end, we aim to measure the semantic similarity between therapist and robot responses to disclosures aligned with the same theme. This will allow us to evaluate whether a robot’s replies in emotionally supportive interactions are aligned in intent and content with those offered by human therapists. Accordingly, we ask-
\textbf{RQ3}: \textit{When people express similar themes of disclosure, to what extent robot responses align semantically with those of human therapists?}


\section{Related Work}


\subsection{Self‑Disclosure and Topic Selection}

Interpersonal‑communication research identify the conditions under which people are most willing to “open up” to one another. For example, Social‑Penetration Theory describes self‑disclosure as a reciprocal process that deepens in breadth and depth as trust accrues over repeated encounters \cite{RefWorks:260}. Nonetheless, it has been previously suggested that when the listener is an artificial agent, a more economic frame would be appropriate \cite{Laban2024SharingFeel}. Social Exchange Theory \cite{Homans1958SocialExchange} holds that people weigh expected rewards (e.g., emotional relief, practical advice) against potential costs (e.g., sharing a personal matter, feeling embarrassment, receiving negative evaluation) \cite{RefWorks:262}. Empirical studies support this notion, showing that listeners who convey warmth and non-judgment through attentive posture or empathic language tend to elicit greater self-disclosure from speakers \cite{Cozby1973Self-disclosure:Review}. Similarly, studies in human–robot interaction (HRI) have found that people tend to disclose more to a robot when they perceive it as more empathetic \cite{Laban2023OpeningBehavior}. This has been also demonstrated with artificial agents that communicate using LLMs, with people disclosing more towards empathetic agents rather than to factual ones \cite{laban_lexi_hai_24}. Moreover, previous studies show that people readily reveal personal information to robots, 
especially when they perceive low judgement and high rapport \cite{Laban2024SharingFeel}. 

When it comes to topic selection, we see a similar trend. Topic selection is a joint coordination problem in conversation, with interlocutors orienting to relevance, face‑management, and their estimate of common ground, with speakers choosing topics whose informational cost matches the dyad’s mutual knowledge \cite{Clark1991GroundingCommunication.}. Large‑scale corpus analyses with the 1,656‑conversation CANDOR corpus, provide empirical support with discussions typically progressing from safe small talk to more personal themes, and the shift is accompanied by a measurable rise in positive affect \cite{Reece2023TheConversation}. Rapid back‑channels (i.e., 
like “mm‑hmm” or head‑nods) and sub‑250ms response latencies, behaviours linked to social connection, peak immediately before such topic changes, suggesting that attentive listening licenses deeper disclosure \cite{Templeton2022FastConversation, Bergey2024FromConversation}. 
Controlled comparisons of self‑disclosure to a human interlocutor, a humanoid robot, and a voice‑only agent found no significant differences in the thematic distribution of topics people chose, implying that the same topic‑selection heuristics generalize to robotic listeners \cite{Chiang2025ComparingAgent}. 



\subsection{Language Alignment in Supportive Dialogue on the Seeker Side}

In therapeutic contexts, linguistic accommodation, including thematic and semantic synchrony between an individual and his counsellor or therapist predict stronger working alliance and superior outcomes \cite{Tickle-Degnen1990}. Person‑centred and other non‑directive models treat topic choice itself as an alliance‑building intervention, inviting clients to lead the agenda while the practitioner signals unconditional positive regard \cite{Rogers1951Client-centeredTheory.}. Conversation‑analytic reviews of dialogical family therapy find that therapists deploy minimal encouragers and selective expansions to keep the floor with the client yet still steer away from institutional or power‑laden topics that could inhibit openness \cite{Ong2023PowerResearch}. Analyses of 106k+ text‑therapy sessions show that exchanges centred on social relationships, emotions, and leisure predict greater client retention, whereas case‑management topics predict dropout, underscoring the motivational value of client‑relevant themes \cite{Imel2024MentalLearning}. In interactions with robots, a previous study found that participants who reported higher levels of loneliness and stress were significantly more likely to engage in socially focused disclosures, such as friendship and connection, while those with lower distress discussed introspective and goal-oriented themes like academic ambition and self-reflection \cite{Laban2025WhatTime}. Similarly, Irfan and Skantze \cite{irfan2025between} showed that people tend to share personal stories with a robot 
in 
emotionally resonant contexts. 

Supportive interactions, whether with humans or robots, are shaped not just by the topics discussed but by how interlocutors align their language and style. Theories such as LSM \cite{Gonzales2010LanguageGroups, Niederhoffer2002LinguisticInteraction, Aafjes-vanDoorn2020LanguageAlliance} and CAT \cite{giles1991contexts} suggest that individuals adapt their linguistic behaviours to match their conversational partners, a process associated with greater rapport and relational satisfaction. Accordingly, individuals seeking support tend to align their language semantics when disclosing about themselves. For example, Yang et al. \cite{10.1145/3290605.3300261} demonstrated that self-disclosure in online cancer support groups is sensitive to channel context. Participants strategically modulated the emotional tone and reciprocity of their messages, illustrating how linguistic alignment and disclosure style are crucial in fostering supportive communication. Yin et al. \cite{Yin2016PrayForDad:Information} demonstrated that the semantics of health-related disclosures on social media, such as the use of social, affective, and temporal language categories, are predictive of whether individuals disclose about themselves or others. This finding reinforces the view that the style and linguistic framing of disclosures are critical components of supportive communication dynamics. Bishop and High \cite{Bishop2023StigmaRelationships} showed how stigma visibility and internalization influence the emotional content, depth, and directness of support-seeking communication, highlighting that supportive interactions are shaped as much by expressive style as by content.

\subsection{Language Alignment in Supportive Dialogue on the Provider Side}

Those providing help and support (e.g., therapists) also align their language towards those seeking help. For example, Borelli et al. \cite{Borelli2019Therapist-clientQuality} provided initial evidence that therapists align their linguistic style with clients during psychotherapy, showing that higher therapist–client language style matching predicts stronger relational attunement and reduced post-treatment distress. Miner et al. \cite{Miner2022AConsistency} applied a computational framework to psychotherapy transcripts and found that therapists adapted their speech features, including pronoun use, emotional polarity, and speech rate, in response to patients’ linguistic behaviour. Notably, therapists often slowed down when patients increased their speech rate, shifted toward present- and future-focused language as sessions progressed, and used more first-person plural pronouns (“we”) over time, indicating dynamic efforts to establish rapport and foster therapeutic alignment. 

Preliminary evidence of AI linguistic alignment during emotional support conversations is still scarce, yet it offers initial support. For example, Tak and Gratch \cite{Tak2023IsEmotion} demonstrated that GPT-4 aligns with human emotional appraisals and affective reasoning across autobiographical and experimental tasks, offering preliminary evidence that LLMs can linguistically reflect human emotional interpretations in emotionally laden situations. Their follow-up work showed that GPT-4’s emotional inferences align more closely with third-person human judgments than with self-assessments \cite{Tak2024GPT-4Perspective}. Sharma et al. \cite{Sharma2020ASupport} introduced the EPITOME framework to model empathy in text-based mental health support and demonstrated that AI systems can detect and enhance empathic language using computational methods, offering initial support for the potential of AI to align linguistically with users during emotionally supportive conversations. Madani et al. \cite{Madani2024SteeringConversations} introduced the Strategy Relevant Attention (SRA) metric to track how well large language models maintain alignment with emotional support strategies during extended conversations, offering early but promising evidence that AI agents can be steered to exhibit consistent supportive behaviour through linguistic and strategic alignment. Brun et al. \cite{Brun2025ExploringAI} demonstrated that even minimal emotional alignment by a GPT-based chatbot, via sentiment-guided prompting, can significantly enhance users’ perceptions of competence, trust, and supportiveness. Another study showed that users interacting with an empathetic LLM-based agent, compared to a neutral counterpart, wrote longer and more positive messages, experienced greater mood improvement, and perceived the agent as more social and emotionally attuned. Interestingly, longer messages provided by the agent were associated with a greater likelihood of users liking the message \cite{laban_lexi_hai_24}. This offers compelling early evidence that both lexical cues, such as message length, and empathy cues contribute to more supportive AI conversations. 

\subsection{Bridging Language Alignment Between Human and Robot Support}

Social robots have demonstrated promising outcomes when delivering psychosocial interventions \cite{RefWorks:404}, including reductions in loneliness \cite{Laban2024BuildingTime,laban_ced_2023,Laban2025WhatTime}, stress \cite{laban_ced_2023,Laban2025WhatTime}, anxiety \cite{Nomura2020,Laban2022SocialTreatment}, and depressive symptoms \cite{chen_jones_2018,Araujo2021EffectsReview}. With the integration of LLMs, recent studies have explored how social robots can move beyond rule-based scripts to offer open-ended, emotionally sensitive dialogue, enabling more natural and supportive interactions \cite{Laban2024SharingFeel,Spitale2024AppropriatenessEvaluation}. These capabilities have been applied to facilitate emotion regulation training \cite{laban2025robot} and deliver positive psychology interventions \cite{SpitaleMicol2025VITA:Coaching}. While these findings are encouraging, most studies have focused on affective or behavioural outcomes rather than examining the conversational structure itself.

Specifically, the field still lacks systematic evidence on whether contemporary robot-generated support mirrors the topical and semantic structure of bona-fide therapy sessions. Although individuals are known to disclose meaningfully to artificial agents under the right conditions \cite{Laban2023OpeningBehavior}, and LLMs can produce language that seems emotionally attuned \cite{Tak2023IsEmotion,Tak2024GPT-4Perspective,Mehra2025BeyondValues}, there is limited research directly comparing how emotional support conversations unfold across human therapists and robots at a content level. This creates a critical gap in our understanding of how AI-powered social robots function as support providers, not just in terms of user outcomes, but in how the themes expressed by users and the responses provided by the agent structurally and semantically align with those in traditional therapy. Prior studies have rarely explored whether people raise similar concerns with robots as they do with therapists, nor whether robot responses reflect the kinds of semantically coherent, empathic replies characteristic of human therapeutic dialogue.

Our study addresses this gap through a comprehensive, data-driven comparison of human-to-human and human-to-robot support conversations. We use clustering methods, LLM-generated theme labelling, and multiple embedding models to examine three key dimensions: \textbf{(1)} whether users disclose similar topics to robots and therapists, \textbf{(2)} whether users express themselves in semantically comparable ways across agent types, and \textbf{(3)} whether robot (LLM-generated) responses semantically resemble those of human therapists when addressing thematically similar disclosures. In doing so, we provide the \textit{first} large-scale semantic alignment analysis between robot-led support and human-delivered therapy, moving beyond behavioural and affective outcomes to explore the substance of supportive language.

\section{Method}
\label{sec:method}

This work includes a secondary analysis of two datasets (see Section \ref{subsec:dataset}): one comprising human-to-human therapist conversations (H2H), and the other involving interactions between a human and a robot providing socio-emotional support (H2R). Rather than conducting high-risk user study prematurely, we adopted a data-driven exploratory approach, analysing two fully anonymized corpora to examine whether the thematic and semantic structures of disclosures align across agent types, thereby offering a first empirical foundation, despite differences in participant demographics and study contexts. Responses from human participants and patients in both datasets were initially clustered and subsequently assigned generated labels and descriptions (see Section \ref{subsec:clustering_pipeling}). For each dataset, we calculated the percentage of responses that could be matched to a cluster derived from the opposing agent type (i.e., robot vs. human therapist). We then conducted a statistical analysis to compare the semantic similarity, using various embedding models, between each H2R unit matched to an H2H cluster and the corresponding H2H units within that cluster, to assess semantic alignment between the two types of agents (see Section \ref{subsec:analysis}). Analysis scripts will be released on OSF upon acceptance for replicability.

\subsection{Datasets}
\label{subsec:dataset}
We used two conversational datasets to analyse potential differences in responses between a human therapist (H2H) and a robot (H2R) operating in an equivalent role. The H2H data involves mid-life adults whereas the H2R data involves university students, however, for the scope of the study we treat thematic and semantic alignment as a content-level phenomenon that is theoretically separable from life-stage-specific concerns in order to test whether robots match humans on the core substance of conversational alignment in support oriented interactions, an essential step before age and context matched clinical trials are feasible (See section \ref{limit}).

\subsubsection{\textbf{Human to Human Dataset}}

The first dataset used was Hugging Face's \textit{NLP Mental Health Conversations}\footnote{Available at: \url{https://huggingface.co/datasets/Amod/mental_health_counseling_conversations}}, compiled from the CounselChat platform. This dataset consists of 3,512 anonymised, text-based counselling interactions between users seeking advice and licensed therapists, who were verified to engage with users in the platform \cite{bertagnolli2020counsel}. Each data instance is structured as a tuple $(x, y)$, where $x$ represents a user's question or concern and $y$ is the corresponding therapist’s written response. 

To build the dataset, the authors filtered and extracted conversations where at least one therapist had responded in detail to a user-submitted question, focusing on exchanges that reflected real-world therapeutic tone, empathy, and relevance. Topics covered include anxiety, depression, trauma, self-esteem, relationships, and more, making the dataset well-suited for analysing linguistic and relational patterns in professional mental health support. This dataset serves as a high-quality resource for modelling human-human therapeutic dialogue, and provides a strong benchmark for evaluating machine-generated interventions in mental health contexts (see examples \cite{Soman2025HumanGeneration,Scotti2023AChatbots,Gollapalli2023IdentifyingTexts}). CounselChat contributors accepted a Creative Commons licence permitting public, anonymised release; no personally identifying information appears in the corpus.

For clarity, disclosure units directed to the therapists will be referred to in this paper as those provided by `\textit{subjects}', while the therapists’ responses will be described as those provided by the `\textit{therapist}'.

\subsubsection{\textbf{Human to Robot Dataset}}
The second dataset used conversational data between humans and a robot (H2R) in supportive interactions, as reported in \cite{laban2025robot}. Data were collected through a five-session robot-led intervention conducted with 21 university students in familiar settings such as university halls and departments (see Figure \ref{fig:set}). Each participant engaged in a structured intervention that contained conversations with the robot ``QTrobot" (LuxAI), which facilitated cognitive reappraisal using a LLM (GPT-3.5), by utilising an open-sourced robotic system \cite{SpitaleMicol2025VITA:Coaching}. The sessions followed the PERMA framework \cite{Seligman2018PERMAWell-being}, with two positive-connotation and one negative-connotation question per session, where the robot guided participants to reinterpret emotionally charged experiences. Data collection included standardized self-report questionnaires as well as interaction logs of the disclosures' content towards the robot, resulting in 560 measurable disclosures by subjects, and 560 measurable responses by the robot. The study was approved by the departmental ethics committee and participants provided informed consent before participating in the study.  For more information about the data collection methodology, see \cite{laban2025robot}.

For clarity, disclosure units directed to the robot will be referred to in this paper as those provided by `\textit{subjects}', while the robot’s responses will be described as those provided by the `\textit{robot}'.

\begin{figure}[h!]
    \centering
    \includegraphics[width=.49\columnwidth]{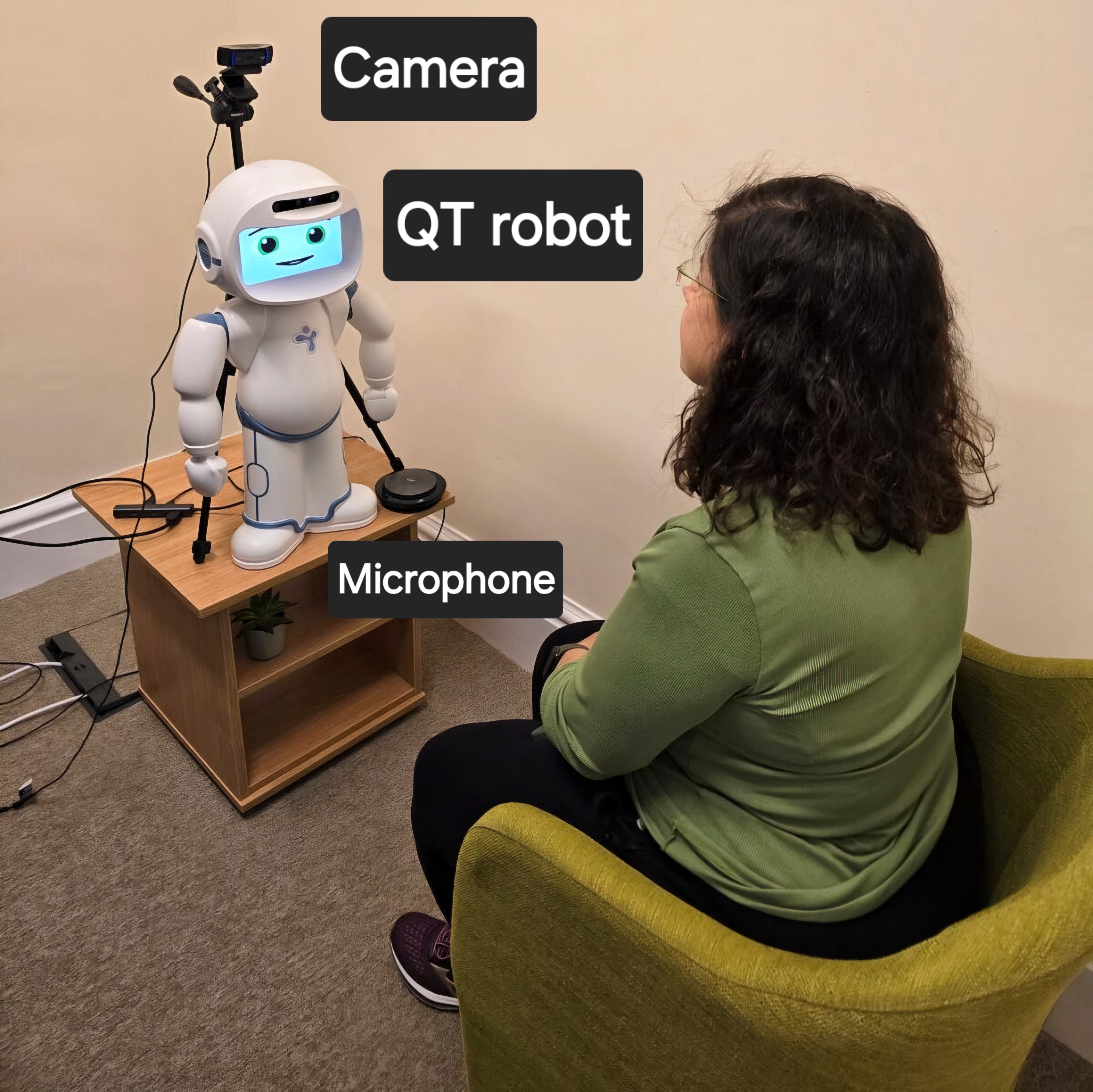} 
    \includegraphics[width=.49\columnwidth]{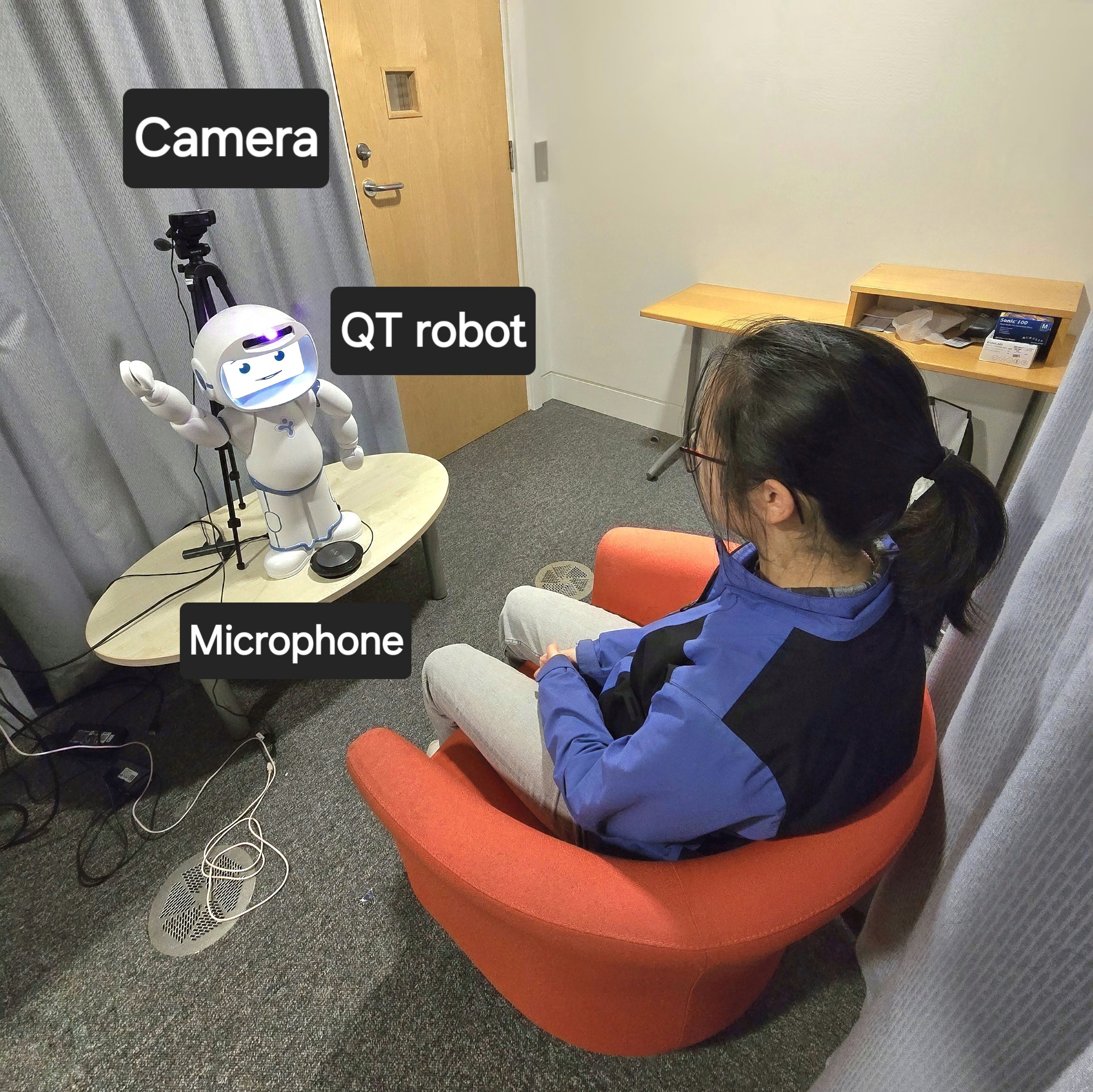}
    \caption{\footnotesize The deployment settings. Image from \cite{laban2025robot}.}
    \label{fig:set}
\end{figure}

\subsection{Preprocessing and Clustering}
\label{subsec:clustering_pipeling}
Both datasets had to be preprocessed before clustering was applied.
Stop words were filtered from the data using the ENGLISH\_STOP\_WORDS list from sklearn \cite{PedregosaFABIANPEDREGOSA2011Scikit-learn:Python} after tokenizing responses by whitespace. Additionally, duplicate responses were removed from each dataset to prevent redundancy.
Each data unit (i.e. a response from the subject) was then converted to word embeddings using the \textit{``all-MiniLM-L6-v2 model"} \cite{10.5555/3495724.3496209} from the SentenceTransformers library \cite{Reimers2019Sentence-BERT:BERT-Networks}.
These embeddings were clustered using the K-means algorithm \cite{likas2003global}, with a fixed random seed to ensure replicability, and the optimal number of clusters was determined using the elbow method \cite{liu2020determine}.
This meant that after K-means was applied, every response was assigned to a specific cluster based on its semantic embedding.
To evaluate internal consistency and spread of each cluster, we computed the Mahalanobis distances, $D_M$, of all embeddings to their respective cluster centroid. We used the inverse of the empirical covariance matrix to account for feature correlation within clusters \cite{de2000mahalanobis}. Mahalanobis distance was selected because, by incorporating the full covariance structure of the sentence-embedding space, it yields a scale-invariant metric that adjusts for correlations between dimensions 
\cite{de2000mahalanobis}.

\begin{equation*}
D_M(x, \mu, \Sigma^{-1}) = \sqrt{(x - \mu)^\top \Sigma^{-1} (x - \mu)}
\end{equation*}
Where:
\begin{itemize}
\item $x \in \mathbb{R}^d \text{: The embedding vector (sample point)}$
\item $\mu \in \mathbb{R}^d \text{: The cluster centroid}$
\item $\Sigma \text{: The covariance matrix of the cluster embeddings}$
\item $d \text{: The embedding dimensionality}$
\end{itemize}

The clusters were explained using GPT-4o-mini \cite{openai2024chatgpt}, which was prompted to provide a label and description for each cluster via the cluster explanation procedure described in \cite{Chiang2025ComparingAgent}. The explanations provided were then validated using the procedure outlined in \cite{Chiang2025ComparingAgent}.

\subsection{Data Analysis}
\label{subsec:analysis}

\subsubsection{Cluster Alignment}
To examine the potential alignment between the two types of agents, we calculated the percentage of responses from each agent type that could be matched to one of the clusters derived from the other agent type.
This process was performed using a distance-based approach, commonly employed in cluster anomaly detection \cite{pu2020hybrid}, to identify responses that could potentially fit into clusters from the opposing conversation type.

To assess whether responses from one agent aligned with the other, we identified the closest centroid for each response, and applied a threshold value to determine fit. If the distance to the nearest centroid exceeded this threshold, we classify the response as not belonging to any cluster.
The threshold, $T_j$, for each cluster $j$ is defined as a scalar multiple $\alpha$ of the average distance from responses within that specific cluster to its centroid, resulting in each cluster having its own distinct threshold based on its internal variability \cite{rehm2007novel}.

\[
T_j = \alpha \cdot \frac{1}{N_j} \sum_{i=1}^{N_j} d_{ij}, \quad \text{where } \mathbf{r}_i \in \text{Cluster } j
\]
Here, \( N_j \) is the total number of responses in cluster \( j \), and \( d_{ij} \) is the distance of response \( \mathbf{r}_i \) to the centroid \( \mathbf{c}_j \).
To assess how closely newly fitted responses aligned with their assigned clusters, we computed the mean Euclidean distance between each response and the centroid of its new respective cluster. Euclidean distance is a widely used metric in clustering analysis for evaluating intra-cluster cohesion and the proximity of data points to cluster centroids \cite{jain1999data}. As a general rule of thumb, any distance above 1.0 suggests increasing deviation from the cluster centre, and any distance from 0.5-1.0 suggests moderate proximity. However, precise thresholds are context dependent and vary based on characteristics of each cluster \cite{jain1999data}.

\subsubsection{Semantic Similarity}
For responses that could be mapped to a cluster of the other agent type, we combined these responses with the already existing responses of said cluster.

For each cluster, we first calculated the mean semantic similarity across all possible pairings between subject responses from the H2R dataset that were mapped to the cluster and all subject responses from the corresponding H2H cluster. This allowed us to assess whether participants expressed semantically similar content when disclosing thematically aligned disclosures to a human therapist and to a robot. For each cluster, we then calculated the mean semantic similarity between responses generated by the robot (for H2R disclosures mapped to that cluster) and all therapist responses in the corresponding H2H cluster. This enabled us to evaluate the extent to which robot responses were semantically comparable to those of human therapists when addressing similarly themed participant disclosures. This resulted in 2,223 comparisons in Cluster 0, 95,816 in Cluster 1, 192 in Cluster 2, 976 in Cluster 3, 844 in Cluster 4, 702 in Cluster 5, and none in Cluster 6. 


We compare this metric across three different embeddings models, including Transformers \cite{10.5555/3295222.3295349} , Word2Vec \cite{Mikolov2013DistributedCompositionality} and BERT \cite{Devlin2018BERT:Understanding}. Each embedding model encodes text in a different way; Sequence Transformers focus more on syntactic structure \cite{selva2021review}, BERT captures contextual meaning of text and Word2Vec provides a baseline for word-level similarity \cite{johnson2024detailed}.
One-sample t-tests were conducted to evaluate whether the semantic similarity scores were significantly greater than a predefined threshold of 0.5, which served as a baseline for similarity above random chance.

\section{Results}
\label{sec:results}

\subsection{Cluster Descriptions and Validation}
\label{subsec:cluster_insights}

\subsubsection{Human to Human Data}
We identified 7 semantically meaningful clusters, with their corresponding within-cluster sum-of-squares scores to be 807.07. This follows values of 770.00, 748.64, 738.86, 720.62, 712.95 and 701.45 for 1 to 6 clusters respectively. This shows an increase in compactness with each additional cluster up to 7.
171 of responses comprised Cluster 0 (17.19\%), 203 in Cluster 1 (20.40\%), 96 in Cluster 2 (9.65\%), 122 in Cluster 3 (12.26\%), 211 in Cluster 4 (21.21\%), 78 in Cluster 5 (7.84\%) and 114 in Cluster 6 (11.46\%).
Table \ref{tab:mahalanobis_h2h} shows Mahalanobis distance statistics for H2H embeddings across all clusters, including mean, standard deviation, minimum and maximum distance. Cluster 1 exhibits the highest average distance to its respective centroid (12.98), suggesting it is spatially dispersed, whereas Clusters 5 shows the tightest distribution (8.19).
Respectively, the LLM-generated cluster labels were (0) Family Conflict and Emotional Stress, (1) Anxieties and Self-Perception Struggles, (2) Struggles with Trust and Toxic Relationships, (3) Emotional Turmoil in Romantic Relationships, (4) Struggles with Mental Health and Isolation, (5) Sexual Confusion and Intimacy Challenges and (6) Trust and Infidelity Issues in Marriage.
Table \ref{tab:h2h_clusters} in the appendix shows cluster labels with their full descriptions.

\begin{table}[ht]
\centering
\caption{\small Mahalanobis Diagnostic Statistics for H2H Clusters}
\label{tab:mahalanobis_h2h}
\begin{tabular}{|c|c|c|c|c|}
\hline
\textbf{Cluster} & \textbf{Mean} & \textbf{Std} & \textbf{Min} & \textbf{Max} \\
\hline
0 & 11.79 & 1.78 & 9.17 & 13.00 \\
1 & 12.98 & 1.89 & 10.00 & 14.18 \\
2 & 8.86  & 1.31 & 6.82 & 9.70  \\
3 & 9.68  & 1.58 & 7.71 & 10.95 \\
4 & 10.04 & 2.01 & 7.20 & 11.46 \\
5 & 8.19  & 1.05 & 6.12 & 8.72  \\
6 & 9.10  & 1.56 & 7.45 & 10.58 \\
\hline
\end{tabular}
\end{table}

\subsubsection{Human to Robot Data}
We identified 6 semantically meaningful clusters which yielded a WCSS value of 393.93. This result followed a decreasing trend in WCSS across cluster counts from one to five, with scores of 444.54, 426.01, 415.50, 404.61, and 397.97, respectively.
113 of responses comprised Cluster 0 (20.18\%), 101 in Cluster 1 (18.04\%), 78 in Cluster 2 (13.93\%), 95 in Cluster 3 (16.96\%), 96 in Cluster 4 (17.14\%) and 77 in Cluster 5 (13.75\%).
Based on the LLM's interpretation, clusters were labeled as (0) Continuous Personal Development and Self-Reflection, (1) Building Connections and Memorable Experiences, (2) Academic Ambition and Future Aspirations, (3) Navigating Interpersonal and Emotional Management, (4) Passion for Learning and Creativity and (5) Friendships: Connection and Loneliness.
Table \ref{tab:h2r_clusters} in the appendix shows cluster labels with their full descriptions.

\subsubsection{Validation of Cluster Explanations} For both datasets, cluster descriptions consistently exhibited the highest semantic similarity to their corresponding centroids (see \cite{Chiang2025ComparingAgent}), indicating that the LLM-generated labels and descriptions effectively captured the core semantics of each cluster (see Table \ref{table:h2h_h2r_validation}).

\begin{table}[ht]
    \centering
    \caption{\small Similarity Between LLM‑Generated Descriptions And Cluster Centroids Across H2H and H2R Data}
    \label{table:h2h_h2r_validation}
    \resizebox{\columnwidth}{!}{%
    \begin{tabular}{|c|c|c|c|c|c|c|c|c|}
        \hline
        \textbf{Data} & \textbf{Clst} & \textbf{Des 0} & \textbf{Des 1} & \textbf{Des 2} & \textbf{Des 3} & \textbf{Des 4} & \textbf{Des 5} & \textbf{Des 6} \\
        \hline
        \multirow{7}{*}{H2H} 
           & 0 & \textbf{0.6141} & 0.3067 & 0.4170 & 0.3511 & 0.3628 & 0.2259 & 0.3948 \\
           & 1 & 0.3508 & \textbf{0.5691} & 0.4091 & 0.3577 & 0.4304 & 0.3781 & 0.3028 \\
           & 2 & 0.4376 & 0.3262 & \textbf{0.5976} & 0.5126 & 0.3260 & 0.3448 & 0.5208 \\
           & 3 & 0.3714 & 0.3624 & 0.5435 & \textbf{0.5636} & 0.3203 & 0.3833 & 0.5325 \\
           & 4 & 0.4449 & 0.5254 & 0.4494 & 0.3595 & \textbf{0.5963} & 0.3572 & 0.3729 \\
           & 5 & 0.2838 & 0.3037 & 0.3724 & 0.3851 & 0.1993 & \textbf{0.6557} & 0.3661 \\
           & 6 & 0.4398 & 0.3016 & 0.5378 & 0.5376 & 0.3118 & 0.4512 & \textbf{0.6766} \\
        \hline
        \multirow{6}{*}{H2R} 
           & 0 & \textbf{0.6088} & 0.4045 & 0.4971 & 0.4319 & 0.6126 & 0.2926 & -- \\
           & 1 & 0.2646 & \textbf{0.5031} & 0.2517 & 0.2243 & 0.3272 & 0.2890 & -- \\
           & 2 & 0.6067 & 0.3395 & \textbf{0.6262} & 0.3456 & 0.5552 & 0.2076 & -- \\
           & 3 & 0.4196 & 0.4181 & 0.2663 & \textbf{0.5816} & 0.4316 & 0.4812 & -- \\
           & 4 & 0.4431 & 0.2762 & 0.3834 & 0.3092 & \textbf{0.5639} & 0.1889 & -- \\
           & 5 & 0.3235 & 0.5918 & 0.2140 & 0.5374 & 0.3523 & \textbf{0.6319} & -- \\
        \hline
    \end{tabular}}
\end{table}

\subsection{Cross-Agent Cluster Alignment}
\label{subsec:cluster_fitting}
With $\alpha=2.0$, 90.88\% of human responses in the H2R dataset fit into the H2H clusters, while 28.04\% of human responses in the H2H dataset fit into the H2R clusters. Based on these results, we combined the H2H clusters with the corresponding H2R responses that fit into each cluster.
From the H2R dataset, 13 of the total responses were able to fit into Cluster 0 of the H2H data (2.56\%). 472 responses were able to fit into Cluster 1 (92.91\%), 2 in Cluster 2 (0.39\%), 8 in Cluster 3 (1.57\%), 4 in Cluster 4 (0.79\%) and 9 in Cluster 5 (1.77\%). No responses were able to fit into Cluster 6 of the H2H data.
We calculated the average Euclidean distance between each response and the centroid of the H2H cluster into which it was classified.
H2R responses mapped to Cluster 0 in the H2H data exhibited an average Euclidean distance of .90 from the cluster centroid (95\% CI [.88, .92]). For the remaining clusters, the average distances were as follows: Cluster 1 – .97 (95\% CI [.97, .98]), Cluster 2 – .80 (95\% CI [.76, .85]), Cluster 3 – .84 (95\% CI [.83, .86]), Cluster 4 – .79 (95\% CI [.72, .87]), and Cluster 5 – .96 (95\% CI [.93, .99]). These distances suggest that while robot-led disclosures generally align with the thematic cores of human therapy clusters, some responses, particularly those related to broader or more introspective themes, tended to occupy more peripheral positions within the cluster space.


\subsection{Semantic Similarity}


\subsubsection{Subject Response Similarity Across Datasets}
For each combined cluster, we calculated the mean pairwise semantic similarity between subject disclosures from the H2H exchange and those from the H2R interaction (see Table \ref{table:participant_similarity}).

\textbf{Transformer:}
The highest similarity found was in Cluster 2 ($M=$ .28, 95\%CI[.26, .29]), and the lowest in Cluster 5 ($M=$ .13, 95\%CI[.12, .14]).
No scores using this model reached significance above the 0.50 threshold ($p > 0.01$ for all) suggesting that semantic consistency was not above chance for this model.

\textbf{Word2Vec:}
Using Word2Vec, scores ranged from ($M=$ .81, 95\%CI[.80, .81]) in Cluster 1, to ($M=$ .90, 95\%CI[.90, .91]) in Cluster 3. There results are consistently high across all clusters and suggest robust semantic alignment between therapist responses to subjects which are above chance.

\textbf{BERT:}
Semantic similarity scores using BERT also indicated high semantic consistency across all clusters, where each exceeded the 0.50 threshold ($p < 0.01$). Values ranged from  ($M=$ .79, 95\%CI[.78, .79]) in Cluster 1 and ($M=$ .82, 95\%CI[.82, .82]) in Clusters 2 and 3. This indicates the semantic alignment are consistent even when taking into account contextual embeddings.

\begin{figure}[h]
    \centering
    \includegraphics[width=\columnwidth]{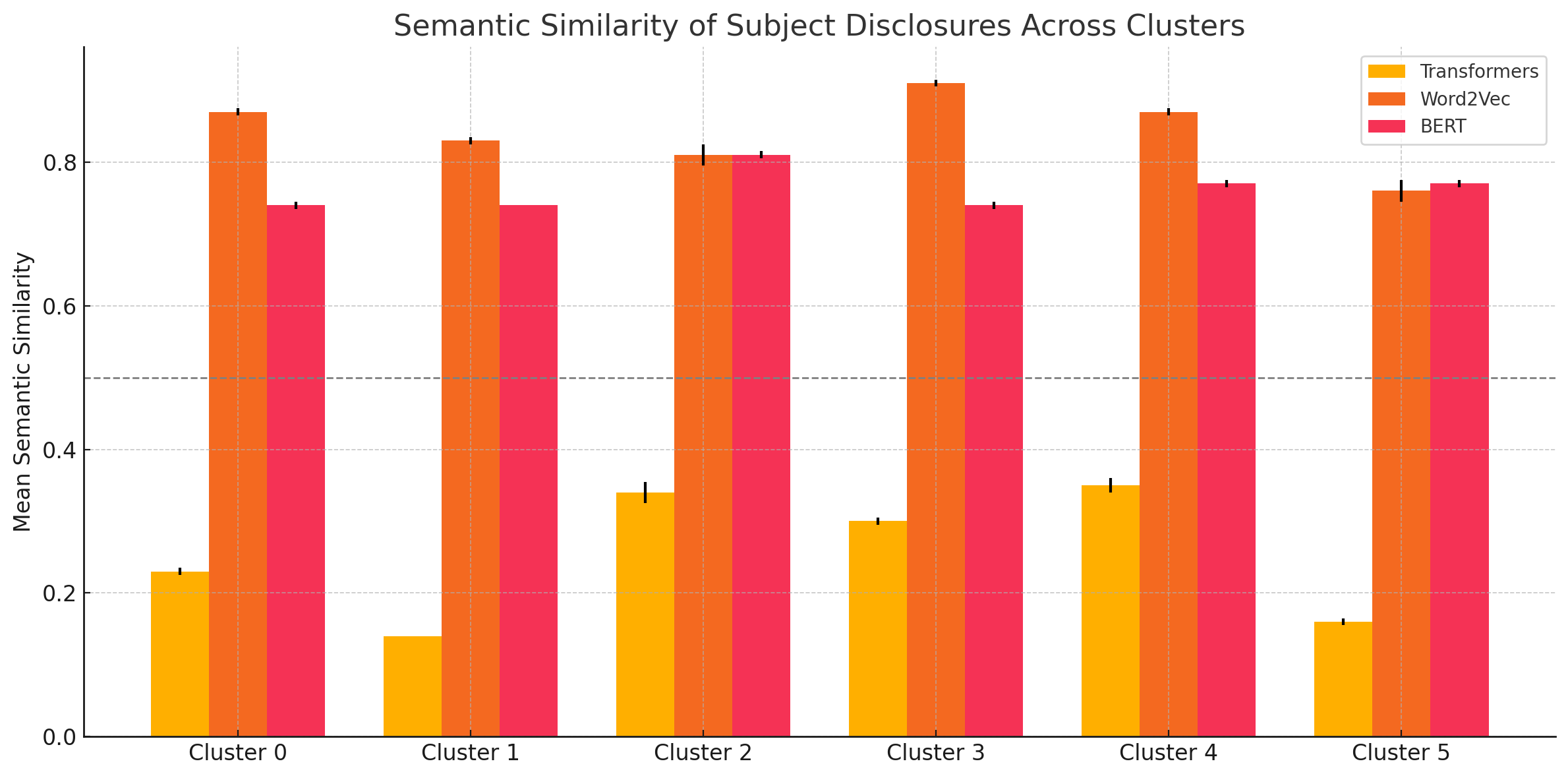} 
    \caption{\footnotesize Mean semantic similarity of subject disclosures across clusters}
    \label{fig:set}
\end{figure}

\begin{table*}[!t]
\centering
\caption{\small Semantic similarity between the subject responses per combined cluster using Sequence Transformers, Word2Vec and BERT as embedding models}
\label{table:participant_similarity}
\resizebox{\textwidth}{!}{
\begin{tabular}{|c|c|c|c|c|c|c|}
\hline
Model & Clst 0 & Clst 1 & Clst 2 & Clst 3 & Clst 4 & Clst 5 \\ 
\hline
Transformers & .23 [.22, .23] & .14 [.14, .14] & .34 [.33, .36] & .30 [.30, .31] & .35 [.34, .36] & .16 [.16, .17] \\ 
Word2Vec     & .87*** [.86, .87] & .83*** [.83, .84] & .81*** [.79, .82] & .91*** [.91, .92] & .87*** [.86, .87] & .76*** [..74, .77] \\ 
BERT         & .74*** [.73, .74] & .74*** [.74, .74] & .81*** [.81, .82] & .74*** [.74, .75] & .77*** [.77, .78] & .77*** [.77, .78] \\ 
\hline
\multicolumn{7}{l}{\textit{Note:} $p < 0.001 = ***$} \\
\end{tabular}}
\end{table*}

\subsubsection{Therapist Response Similarity Across Datasets}
For each combined cluster, we calculated the mean pairwise semantic similarity between the human therapist responses from the H2H exchange and the robot response from the H2R interaction (see Table \ref{table:participant_similarity}).

\textbf{Transformer:}
The highest similarity found was in Cluster 4 ($M=$ .35, 95\%CI[.34, .36]), and the lowest in Cluster 1 ($M=$ .14, 95\%CI[.14, .14]).
No scores using this model reached significance above the 0.50 threshold ($p > 0.01$ for all) suggesting that semantic consistency was not above chance for this model.

\textbf{Word2Vec:}
Using Word2Vec, scores ranged from ($M=$ .81, 95\%CI[.79, .82]) in Cluster 2, to ($M=$ .91, 95\%CI[.91, .92]) in Cluster 3. There results are consistently high across all clusters and suggest robust semantic alignment between therapist responses to subjects which are above chance.

\textbf{BERT:}
Semantic similarity scores using BERT also indicated high semantic consistency across all clusters, where each exceeded the 0.50 threshold ($p < 0.01$). Values ranged from  ($M=$ .74, 95\%CI[.73, .74]) in Clusters 0,1 and 3, to ($M=$ .81, 95\%CI[.81, .82]) in Cluster 2. This indicates the semantic alignment are consistent even when taking into account contextual embeddings.

\begin{figure}[h]
    \centering
    \includegraphics[width=\columnwidth]{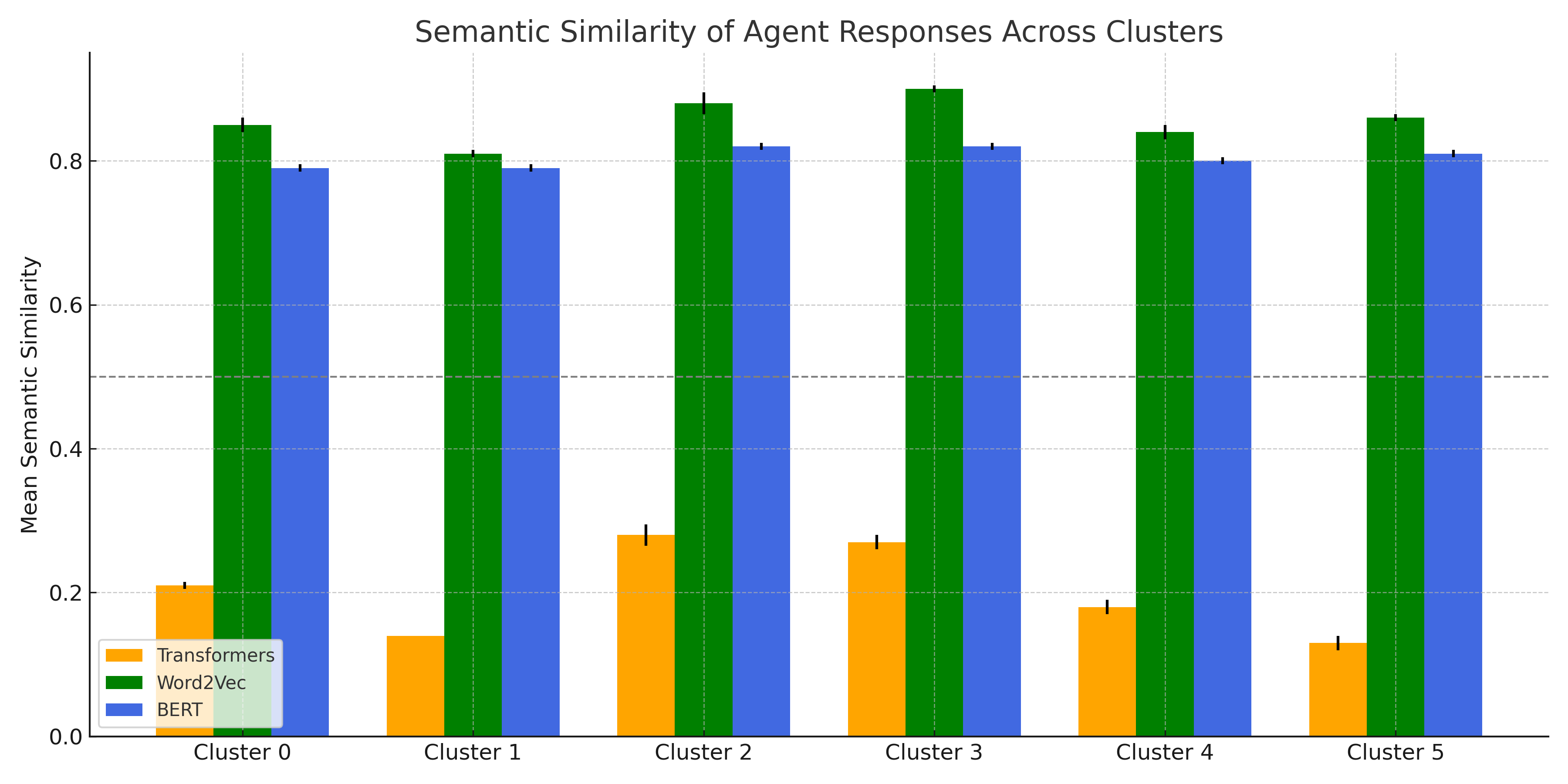} 
    \caption{\footnotesize Mean semantic similarity of agent responses across clusters}
    \label{fig:set}
\end{figure}

\begin{table*}[!t]
\centering
\caption{\small Semantic similarity between the robot and human therapist responses per combined cluster using Sequence Transformers, Word2Vec and BERT as embedding models}
\label{table:therapist_similarity}
\resizebox{\textwidth}{!}{
\begin{tabular}{|c|c|c|c|c|c|c|}
\hline
Model & Clst 0 & Clst 1 & Clst 2 & Clst 3 & Clst 4 & Clst 5 \\ 
\hline
Transformers & .21 [.20, .21] & .14 [.14, .14] & .28 [.26, .29] & .27 [.26, .28] & .18 [.17, .19] & .13 [.12, .14] \\ 
Word2Vec     & .85*** [.84, .86] & .81*** [.80, .81] & .88*** [.86, .89] & 0.90*** [.90, .91] & .84*** [.83, .85] & .86*** [..86, .87] \\ 
BERT         & .79*** [.79, .80] & .79*** [.78, .79] & .82*** [.82, .83] & .82*** [.82, .83] & .80*** [..80, .81] & .81*** [.81, .82] \\ 
\hline
\multicolumn{7}{l}{\textit{Note:} $p < 0.001 = ***$} \\
\end{tabular}}
\end{table*}

\section{Discussion}
\label{sec:discussion}

This study set out to examine whether robot-led supportive conversations can mirror the semantic and thematic structure of traditional therapeutic dialogue. Through cross-agent comparison of participant disclosures and corresponding agent responses, we aimed to address three central questions: whether topics shared with a robot align with those shared with human therapists \textbf{(RQ1)}, whether users articulate disclosures in semantically similar ways across agents \textbf{(RQ2)}, and whether robots respond in ways that semantically resemble human therapists \textbf{(RQ3)}.

\subsection{Aligning Disclosure Themes Across Agents}

Our findings demonstrate a high degree of thematic overlap between the human-to-robot (H2R) and human-to-human (H2H) conversations (RQ1). Remarkably, over 90\% of participant disclosures made to the robot could be mapped to clusters derived from human therapy sessions. This suggests that participants bring similarly structured concerns, such as trust, mental health, interpersonal difficulties, and self-perception struggles, regardless of whether they are speaking to a robot or a human therapist. These results support previous work suggesting that individuals are willing to disclose emotionally meaningful content to artificial agents, especially when those interactions are perceived as non-judgmental and private \cite{Laban2023OpeningBehavior, Laban2024BuildingTime, laban_ced_2023, laban2025robot,Laban2024SharingFeel}. However, this alignment was not entirely symmetric: only 28\% of human therapy disclosures mapped back onto robot-derived clusters. This asymmetry could stem from the structured nature of the robot-led intervention, which may have guided participants toward specific themes (e.g., personal growth, academic ambition, and social experiences) tied to the intervention context. In contrast, the H2H data reflected a broader range of organically elicited therapeutic concerns. This distinction underscores the importance of contextual framing and interactional goals in shaping the thematic landscape of emotional disclosures.

Moreover, it is important to consider that over 90\% of the units in the H2R were loaded into cluster 1 of the H2H dataset. A potential reason for why most of the robot‑led disclosures gravitated toward Cluster 1 (Anxieties and Self‑Perception Struggles) could be related to the method, the intervention design, as well as to psychological tendencies in HRI. Cluster 1 is the broadest and most internally dispersed region in the H2H embedding space (highest Mahalanobis distance), so the distance‑based fitting rule granted it the largest “capture radius”, statistically predisposing disclosures to land there \cite{Gallego2013OnDistributions}. Conceptually, the intervention paradigm that guided the robot repeatedly invited participants to reflect on positive and negative feelings \cite{laban2025robot} for simulating cognitive reappraisals, prompts that naturally elicit self‑evaluative language which sit at the heart of Cluster 1. Coupled with the developmental profile of our young adults sample, whose primary concerns centre on identity formation and peer acceptance \cite{Rageliene2016LinksReview}, the intervention funnelled the majority of utterances toward this thematic basin. Moreover, previous studies show that people tend to engage in affective labelling when interacting with robots \cite{Laban2024BuildingTime, laban_ced_2023}, often using the robot’s nonjudgemental nature as an opportunity for introspection \cite{Laban2024SharingFeel}, as evident from participants’ qualitative responses (see \cite{laban2025robot}). Hence, the dominance of Cluster 1 reflects both a genuine topical convergence (what participants chose to discuss when prompted) and an artefact of the clustering geometry. Recognising this dual driver clarifies why other, more specific themes remained under‑represented and suggests that future robot programmes may need targeted follow‑up probes or adaptive branching to diversify the thematic space they elicit.

\subsection{Semantic Similarity in Participant Disclosures}

In addressing RQ2, we observed high semantic similarity in participants’ disclosures across agent types when using Word2Vec and BERT embeddings, though not with Sequence Transformers. These results suggest that when discussing similar themes, people tend to convey comparable semantic content, regardless of whether they are interacting with a robot or a human therapist. This finding adds to a growing body of evidence suggesting that users adapt their linguistic behaviour to the social context of the interaction \cite{Gonzales2010LanguageGroups,Ireland2011LanguageStability,Niederhoffer2002LinguisticInteraction}, and further demonstrates that robot-led interactions can elicit natural, semantically rich expressions of emotional content. The results also align with theoretical accounts such as CAT \cite{giles1991contexts} and LSM \cite{Gonzales2010LanguageGroups,Aafjes-vanDoorn2020LanguageAlliance}, which propose that interlocutors tend to align their speech patterns with those of their conversation partners. While our study did not explicitly test for linguistic convergence, the high semantic similarity indicates that people may be internalizing the robot’s perceived social role as a support provider and engaging accordingly.

\subsection{Semantic Similarity in Agent Responses}

When examining the responses generated by the robot and the human therapist (RQ3), we again found robust semantic alignment using BERT and Word2Vec embeddings. These models revealed consistently high similarity scores across clusters, suggesting that when given disclosures of similar semantic content, the robot produced responses that matched the intent and contextual depth of human therapists. This outcome speaks to the sophistication of current LLM-powered agents in generating meaningful, contextually appropriate replies. However, it is important to acknowledge that these results were model-dependent, as Transformer-based embeddings did not yield semantic similarity scores above chance, indicating that surface-level linguistic structures alone may not fully capture the nuanced quality of these interactions \cite{selva2021review}. These findings contribute to our understanding of robot-mediated care and emotional support. Rather than attempting to mimic human therapy, robot-led support may be seen as offering a distinct but semantically grounded mode of engagement, capable of addressing many of the same concerns and providing emotionally attuned responses, albeit within a more constrained interactional scope.

\subsection{Implications, Limitations and Future Directions}
\label{limit}
Our results have both theoretical and practical implications. First, they support the notion that robot-led interactions can be meaningfully situated within frameworks of therapeutic dialogue, at least in terms of topic alignment and response quality. Second, they offer empirical grounding for the deployment of social robots in settings where human mental health resources are limited, suggesting that such systems could augment existing support structures by offering preliminary or complementary assistance.

Nevertheless, limitations remain. The robot-led conversations were conducted in a structured intervention context, which may not generalize to more open-ended therapeutic scenarios. Nevertheless, limitations remain. The robot-led conversations were conducted in a structured intervention context, which may be limited to generalise to more open-ended therapeutic scenarios. Moreover, the H2R corpus involves university students, whereas the H2H corpus represents mid-life adults disclosing to a licensed clinician. A perfectly controlled comparison, either pairing a similar student population with a human therapist or recruiting mid-life adults for robot-mediated sessions that allow discussion of higher-risk topics, would remove this demographic mismatch. Yet such trials are currently constrained by non-trivial ethical and logistical barriers, such as therapist-licensure rules, institutional duty-of-care policies, and the absence of crisis-response protocols for autonomous agents, making a live, high-risk robot-therapy study ethically and logistically premature. While employing such perfect comparisons is methodologically desirable, such studies are currently absent in HRI due to this prematurity \cite{Laban2025CriticalWellbeing}. Consequently, we adopted a data-driven exploratory approach by analysing two available, fully anonymised corpora, in order to test whether the relative thematic and semantic structures of disclosures align across agent types, obtaining the first empirical picture of commonalities and divergences across agent types in such unique support settings before deploying such high-risk user studies. At this point in time, considering the maturity of the field, data-driven studies are needed to further provide the insights required to fully realise the potential of socially assistive robots and to prepare for clinically valid randomised controlled trials. Mapping these commonalities and divergences supplies the empirical groundwork and concrete topic targets that future, population-matched, clinically supervised studies will need before proceeding responsibly.

Additionally, while semantic similarity offers an important proxy for alignment, it does not capture the full spectrum of therapeutic efficacy, such as emotional attunement, long-term behavioural change, or perceived empathy. Future research should explore how these semantic similarities translate into user outcomes, such as emotional relief, self-insight, or trust. It would also be valuable to investigate how disclosure dynamics evolve in less structured interactions and across longer time scales (e.g., \cite{Laban2024BuildingTime,laban_ced_2023}). Finally, incorporating measures of alignment beyond semantics, such as turn-taking behaviour (e.g., \cite{Skantze2021Turn-takingReview,10.5555/3721488.3721593}), prosody (e.g., \cite{Laban2021}), behaviour (e.g., \cite{Perez-Osorio2015GazeGoals, Perez-Osorio2017ExpectationsEffect, laban2025robot}) or physiological synchrony (e.g., \cite{Riddoch2023InvestigatingRobot,Perez-Osorio2017ExpectationsEffect}), could provide a more complete understanding of the interactional alignment in robot-led support.

\section{Conclusion}

This study provides the first large-scale, systematic comparison of thematic and semantic alignment between human therapists and robot-led supportive conversations. Our results show that participants disclose similarly structured concerns to both agents and express themselves in semantically consistent ways, especially when analysed with contextual embeddings. Most disclosures made to the robot were successfully mapped onto themes commonly addressed in therapy, such as anxieties, self-perception struggles, and interpersonal challenges, demonstrating notable overlap in both content and concern. Participants also articulated their concerns in semantically similar ways across human and robot interactions, suggesting that the framing of disclosures remains consistent regardless of the support agent. However, we also observed asymmetries in theme distribution and cluster specificity, with certain more complex or sensitive therapeutic themes appearing less frequently in robot-led conversations. Moreover, robot responses powered by LLMs exhibited strong semantic alignment with human therapist replies, suggesting that robot-led support can reflect the core content and intent of traditional therapy dialogue. While differences in interaction context and thematic diversity remain, our findings underscore the potential of LLM-empowered robots to meaningfully participate in emotionally supportive communication. Future work should explore how such semantic alignments relate to user outcomes and long-term relational dynamics.

While our findings highlight the semantic coherence and thematic alignment between robot, and human-led, supportive conversations, it is essential to emphasize that these results should not be interpreted as evidence that robots can or should replace human therapists. Therapy is a deeply relational and context-sensitive process that draws on embodied presence, emotional nuance, and clinical judgment, capacities \cite{Rogers1951Client-centeredTheory.} that remain uniquely human. However, the alignment observed in this study helps us understand why individuals may be drawn to robot-led supportive interactions, particularly what they would like to share with robots and to what extent it is aligned with the topic for which people turn to therapists, and the semantic capacity of robots to respond appropriately. Understanding these dynamics allows us to better grasp the role robots might play, not as substitutes, but as complementary tools that can extend access to support, especially in resource-constrained settings or early stages of help-seeking. As such, ethical deployment must prioritize transparency, user agency, and integration with professional care pathways.

\section{Acknowledgments}

G. Laban and H. Gunes are supported by the EPSRC project ARoEQ under grant ref. EP/R030782/1. S. Chiang undertook this work while she was a visiting undergraduate student at the AFAR Lab, Department of CST, University of Cambridge.

\bibliographystyle{ACM-Reference-Format}
\balance\bibliography{ref2}


\begin{thebibliography}{79}


\ifx \showCODEN    \undefined \def \showCODEN     #1{\unskip}     \fi
\ifx \showISBNx    \undefined \def \showISBNx     #1{\unskip}     \fi
\ifx \showISBNxiii \undefined \def \showISBNxiii  #1{\unskip}     \fi
\ifx \showISSN     \undefined \def \showISSN      #1{\unskip}     \fi
\ifx \showLCCN     \undefined \def \showLCCN      #1{\unskip}     \fi
\ifx \shownote     \undefined \def \shownote      #1{#1}          \fi
\ifx \showarticletitle \undefined \def \showarticletitle #1{#1}   \fi
\ifx \showURL      \undefined \def \showURL       {\relax}        \fi
\providecommand\bibfield[2]{#2}
\providecommand\bibinfo[2]{#2}
\providecommand\natexlab[1]{#1}
\providecommand\showeprint[2][]{arXiv:#2}

\bibitem[Aafjes-van Doorn et~al\mbox{.}(2020)]%
        {Aafjes-vanDoorn2020LanguageAlliance}
\bibfield{author}{\bibinfo{person}{Katie Aafjes-van Doorn}, \bibinfo{person}{John Porcerelli}, {and} \bibinfo{person}{Lena~Christine M{\"{u}}ller-Frommeyer}.} \bibinfo{year}{2020}\natexlab{}.
\newblock \showarticletitle{{Language style matching in psychotherapy: An implicit aspect of alliance}}.
\newblock \bibinfo{journal}{\emph{Journal of Counseling Psychology}} \bibinfo{volume}{67}, \bibinfo{number}{4} (\bibinfo{date}{7} \bibinfo{year}{2020}), \bibinfo{pages}{509--522}.
\newblock
\showISSN{00220167}
\href{https://doi.org/10.1037/COU0000433}{doi:\nolinkurl{10.1037/COU0000433}}


\bibitem[Altman and Taylor(1973)]%
        {RefWorks:260}
\bibfield{author}{\bibinfo{person}{Irwin Altman} {and} \bibinfo{person}{Dalmas~A Taylor}.} \bibinfo{year}{1973}\natexlab{}.
\newblock \bibinfo{booktitle}{\emph{{Social penetration: The development of interpersonal relationships}}}.
\newblock \bibinfo{publisher}{Holt, Rinehart {\&} Winston}, \bibinfo{address}{Oxford, England}. viii, 212--viii, 212 pages.
\newblock


\bibitem[Araujo et~al\mbox{.}(2021)]%
        {Araujo2021EffectsReview}
\bibfield{author}{\bibinfo{person}{Bruno Sanchez~de Araujo}, \bibinfo{person}{Marcelo Fantinato}, \bibinfo{person}{Sarajane Marques~Peres}, \bibinfo{person}{Ruth Caldeira~de Melo}, \bibinfo{person}{Samila Sathler~Tavares Batistoni}, \bibinfo{person}{Meire Cachioni}, {and} \bibinfo{person}{Patrick~C.K. Hung}.} \bibinfo{year}{2021}\natexlab{}.
\newblock \showarticletitle{{Effects of social robots on depressive symptoms in older adults: a scoping review}}.
\newblock \bibinfo{journal}{\emph{Library Hi Tech}} (\bibinfo{year}{2021}).
\newblock
\showISSN{07378831}
\href{https://doi.org/10.1108/LHT-09-2020-0244/FULL/PDF}{doi:\nolinkurl{10.1108/LHT-09-2020-0244/FULL/PDF}}


\bibitem[Bergey and DeDeo(2024)]%
        {Bergey2024FromConversation}
\bibfield{author}{\bibinfo{person}{Claire~Augusta Bergey} {and} \bibinfo{person}{Simon DeDeo}.} \bibinfo{year}{2024}\natexlab{}.
\newblock \showarticletitle{{From "um" to "yeah": Producing, predicting, and regulating information flow in human conversation}}.
\newblock  (\bibinfo{date}{3} \bibinfo{year}{2024}).
\newblock
\urldef\tempurl%
\url{https://arxiv.org/abs/2403.08890v1}
\showURL{%
\tempurl}


\bibitem[Bertagnolli(2020)]%
        {bertagnolli2020counsel}
\bibfield{author}{\bibinfo{person}{Nicolas Bertagnolli}.} \bibinfo{year}{2020}\natexlab{}.
\newblock \bibinfo{title}{Counsel chat: Bootstrapping high-quality therapy data}.
\newblock
\urldef\tempurl%
\url{https://medium.com/data-science/counsel-chat-bootstrapping-high-quality-therapy-data-971b419f33da}
\showURL{%
\tempurl}


\bibitem[Bishop and High(2023)]%
        {Bishop2023StigmaRelationships}
\bibfield{author}{\bibinfo{person}{Rachael~E. Bishop} {and} \bibinfo{person}{Andrew~C. High}.} \bibinfo{year}{2023}\natexlab{}.
\newblock \showarticletitle{{Stigma and Supportive Communication in the Context of Mental or Emotional Distress: An Extension of the Paradox of Support Seeking in Close Relationships}}.
\newblock \bibinfo{journal}{\emph{Communication Research}} (\bibinfo{year}{2023}).
\newblock
\showISSN{15523810}
\href{https://doi.org/10.1177/00936502231189811}{doi:\nolinkurl{10.1177/00936502231189811}}


\bibitem[Borelli et~al\mbox{.}(2019)]%
        {Borelli2019Therapist-clientQuality}
\bibfield{author}{\bibinfo{person}{Jessica~L. Borelli}, \bibinfo{person}{Lucas Sohn}, \bibinfo{person}{Binghuang~A. Wang}, \bibinfo{person}{Kajung Hong}, \bibinfo{person}{Cindy Decoste}, {and} \bibinfo{person}{Nancy~E. Suchman}.} \bibinfo{year}{2019}\natexlab{}.
\newblock \showarticletitle{{Therapist-client language matching: Initial promise as a measure of therapist-client relationship quality}}.
\newblock \bibinfo{journal}{\emph{Psychoanalytic Psychology}} \bibinfo{volume}{36}, \bibinfo{number}{1} (\bibinfo{date}{1} \bibinfo{year}{2019}), \bibinfo{pages}{9--18}.
\newblock
\showISSN{19391331}
\href{https://doi.org/10.1037/PAP0000177}{doi:\nolinkurl{10.1037/PAP0000177}}


\bibitem[Brun et~al\mbox{.}(2025)]%
        {Brun2025ExploringAI}
\bibfield{author}{\bibinfo{person}{Antonin Brun}, \bibinfo{person}{Ruying Liu}, \bibinfo{person}{Aryan Shukla}, \bibinfo{person}{Frances Watson}, {and} \bibinfo{person}{Jonathan Gratch}.} \bibinfo{year}{2025}\natexlab{}.
\newblock \showarticletitle{{Exploring Emotion-Sensitive LLM-Based Conversational AI}}.
\newblock  (\bibinfo{date}{2} \bibinfo{year}{2025}).
\newblock
\urldef\tempurl%
\url{https://arxiv.org/abs/2502.08920v1}
\showURL{%
\tempurl}


\bibitem[Chen et~al\mbox{.}(2018)]%
        {chen_jones_2018}
\bibfield{author}{\bibinfo{person}{Shu~Chuan Chen}, \bibinfo{person}{Cindy Jones}, {and} \bibinfo{person}{Wendy Moyle}.} \bibinfo{year}{2018}\natexlab{}.
\newblock \showarticletitle{{Social Robots for Depression in Older Adults: A Systematic Review}}.
\newblock \bibinfo{journal}{\emph{Journal of Nursing Scholarship}} \bibinfo{volume}{50}, \bibinfo{number}{6} (\bibinfo{date}{11} \bibinfo{year}{2018}), \bibinfo{pages}{612--622}.
\newblock
\showISSN{1547-5069}
\href{https://doi.org/10.1111/JNU.12423}{doi:\nolinkurl{10.1111/JNU.12423}}


\bibitem[Chiang et~al\mbox{.}(2025)]%
        {Chiang2025ComparingAgent}
\bibfield{author}{\bibinfo{person}{Sophie Chiang}, \bibinfo{person}{Guy Laban}, \bibinfo{person}{Emily~S. Cross}, {and} \bibinfo{person}{Hatice Gunes}.} \bibinfo{year}{2025}\natexlab{}.
\newblock \showarticletitle{{Comparing Self-Disclosure Themes and Semantics to a Human, a Robot, and a Disembodied Agent}}. In \bibinfo{booktitle}{\emph{2025 34nd IEEE International Conference on Robot and Human Interactive Communication (RO-MAN)}}.
\newblock


\bibitem[Clark and Brennan(1991)]%
        {Clark1991GroundingCommunication.}
\bibfield{author}{\bibinfo{person}{Herbert~H. Clark} {and} \bibinfo{person}{Susan~E. Brennan}.} \bibinfo{year}{1991}\natexlab{}.
\newblock \showarticletitle{{Grounding in communication.}}
\newblock \bibinfo{journal}{\emph{Perspectives on socially shared cognition.}} (\bibinfo{date}{10} \bibinfo{year}{1991}), \bibinfo{pages}{127--149}.
\newblock
\href{https://doi.org/10.1037/10096-006}{doi:\nolinkurl{10.1037/10096-006}}


\bibitem[Cozby(1973)]%
        {Cozby1973Self-disclosure:Review}
\bibfield{author}{\bibinfo{person}{Paul~C. Cozby}.} \bibinfo{year}{1973}\natexlab{}.
\newblock \showarticletitle{{Self-disclosure: A literature review}}.
\newblock \bibinfo{journal}{\emph{Psychological Bulletin}} \bibinfo{volume}{79}, \bibinfo{number}{2} (\bibinfo{date}{2} \bibinfo{year}{1973}), \bibinfo{pages}{73--91}.
\newblock
\showISSN{00332909}
\href{https://doi.org/10.1037/H0033950}{doi:\nolinkurl{10.1037/H0033950}}


\bibitem[De~Maesschalck et~al\mbox{.}(2000)]%
        {de2000mahalanobis}
\bibfield{author}{\bibinfo{person}{Roy De~Maesschalck}, \bibinfo{person}{Delphine Jouan-Rimbaud}, {and} \bibinfo{person}{D{\'e}sir{\'e}~L Massart}.} \bibinfo{year}{2000}\natexlab{}.
\newblock \showarticletitle{The mahalanobis distance}.
\newblock \bibinfo{journal}{\emph{Chemometrics and intelligent laboratory systems}} \bibinfo{volume}{50}, \bibinfo{number}{1} (\bibinfo{year}{2000}), \bibinfo{pages}{1--18}.
\newblock


\bibitem[Devlin et~al\mbox{.}(2018)]%
        {Devlin2018BERT:Understanding}
\bibfield{author}{\bibinfo{person}{Jacob Devlin}, \bibinfo{person}{Ming~Wei Chang}, \bibinfo{person}{Kenton Lee}, {and} \bibinfo{person}{Kristina Toutanova}.} \bibinfo{year}{2018}\natexlab{}.
\newblock \showarticletitle{{BERT: Pre-training of Deep Bidirectional Transformers for Language Understanding}}.
\newblock \bibinfo{journal}{\emph{NAACL HLT 2019}}  \bibinfo{volume}{1} (\bibinfo{date}{10} \bibinfo{year}{2018}), \bibinfo{pages}{4171--4186}.
\newblock
\showISBNx{9781950737130}
\urldef\tempurl%
\url{https://arxiv.org/abs/1810.04805v2}
\showURL{%
\tempurl}


\bibitem[Gallego et~al\mbox{.}(2013)]%
        {Gallego2013OnDistributions}
\bibfield{author}{\bibinfo{person}{Guillermo Gallego}, \bibinfo{person}{Carlos Cuevas}, \bibinfo{person}{Raúl Mohedano}, {and} \bibinfo{person}{Narciso Garc{\'{i}}a}.} \bibinfo{year}{2013}\natexlab{}.
\newblock \showarticletitle{{On the mahalanobis distance classification criterion for multidimensional normal distributions}}.
\newblock \bibinfo{journal}{\emph{IEEE Transactions on Signal Processing}} \bibinfo{volume}{61}, \bibinfo{number}{17} (\bibinfo{date}{9} \bibinfo{year}{2013}), \bibinfo{pages}{4387--4396}.
\newblock
\showISSN{1053587X}
\href{https://doi.org/10.1109/TSP.2013.2269047}{doi:\nolinkurl{10.1109/TSP.2013.2269047}}


\bibitem[Gasiorek et~al\mbox{.}(2021)]%
        {Gasiorek2021InteractionalPsychology}
\bibfield{author}{\bibinfo{person}{Jessica Gasiorek}, \bibinfo{person}{Ann Weatherall}, {and} \bibinfo{person}{Bernadette Watson}.} \bibinfo{year}{2021}\natexlab{}.
\newblock \showarticletitle{{Interactional Adjustment: Three Approaches in Language and Social Psychology}}.
\newblock \bibinfo{journal}{\emph{Journal of Language and Social Psychology}} \bibinfo{volume}{40}, \bibinfo{number}{1} (\bibinfo{date}{1} \bibinfo{year}{2021}), \bibinfo{pages}{102--119}.
\newblock
\showISSN{15526526}
\href{https://doi.org/10.1177/0261927X20965652}{doi:\nolinkurl{10.1177/0261927X20965652}}


\bibitem[Gelso and Carter(1985)]%
        {Gelso1985ThePsychotherapy}
\bibfield{author}{\bibinfo{person}{Charles~J. Gelso} {and} \bibinfo{person}{Jean~A. Carter}.} \bibinfo{year}{1985}\natexlab{}.
\newblock \showarticletitle{{The Relationship in Counseling and Psychotherapy}}.
\newblock \bibinfo{journal}{\emph{The Counseling Psychologist}} \bibinfo{volume}{13}, \bibinfo{number}{2} (\bibinfo{year}{1985}), \bibinfo{pages}{155--243}.
\newblock
\showISSN{15523861}
\href{https://doi.org/10.1177/0011000085132001}{doi:\nolinkurl{10.1177/0011000085132001}}


\bibitem[Giles et~al\mbox{.}(1991)]%
        {giles1991contexts}
\bibfield{author}{\bibinfo{person}{Howard Giles}, \bibinfo{person}{Justine Coupland}, {and} \bibinfo{person}{Nikolas Coupland}.} \bibinfo{year}{1991}\natexlab{}.
\newblock \bibinfo{booktitle}{\emph{Contexts of accommodation: Developments in applied sociolinguistics}}.
\newblock \bibinfo{publisher}{Cambridge University Press}.
\newblock


\bibitem[Gollapalli et~al\mbox{.}(2023)]%
        {Gollapalli2023IdentifyingTexts}
\bibfield{author}{\bibinfo{person}{Sujatha~Das Gollapalli}, \bibinfo{person}{Beng~Heng Ang}, {and} \bibinfo{person}{See~Kiong Ng}.} \bibinfo{year}{2023}\natexlab{}.
\newblock \showarticletitle{{Identifying Early Maladaptive Schemas from Mental Health Question Texts}}.
\newblock \bibinfo{journal}{\emph{Findings of the Association for Computational Linguistics: EMNLP 2023}} (\bibinfo{year}{2023}), \bibinfo{pages}{11832--11843}.
\newblock
\showISBNx{9798891760615}
\href{https://doi.org/10.18653/V1/2023.FINDINGS-EMNLP.792}{doi:\nolinkurl{10.18653/V1/2023.FINDINGS-EMNLP.792}}


\bibitem[Gonzales et~al\mbox{.}(2010)]%
        {Gonzales2010LanguageGroups}
\bibfield{author}{\bibinfo{person}{Amy~L. Gonzales}, \bibinfo{person}{Jeffrey~T. Hancock}, {and} \bibinfo{person}{James~W. Pennebaker}.} \bibinfo{year}{2010}\natexlab{}.
\newblock \showarticletitle{{Language Style Matching as a Predictor of Social Dynamics in Small Groups}}.
\newblock \bibinfo{journal}{\emph{Communication Research}} \bibinfo{volume}{37}, \bibinfo{number}{1} (\bibinfo{date}{2} \bibinfo{year}{2010}), \bibinfo{pages}{3--19}.
\newblock
\showISSN{00936502}
\href{https://doi.org/10.1177/0093650209351468}{doi:\nolinkurl{10.1177/0093650209351468}}


\bibitem[Henschel et~al\mbox{.}(2021)]%
        {Henschel2021}
\bibfield{author}{\bibinfo{person}{Anna Henschel}, \bibinfo{person}{Guy Laban}, {and} \bibinfo{person}{Emily~S Cross}.} \bibinfo{year}{2021}\natexlab{}.
\newblock \showarticletitle{{What Makes a Robot Social? A Review of Social Robots from Science Fiction to a Home or Hospital Near You}}.
\newblock \bibinfo{journal}{\emph{Current Robotics Reports}} \bibinfo{number}{2} (\bibinfo{year}{2021}), \bibinfo{pages}{9--19}.
\newblock
\showISSN{2662-4087}
\href{https://doi.org/10.1007/s43154-020-00035-0}{doi:\nolinkurl{10.1007/s43154-020-00035-0}}


\bibitem[Homans(1958)]%
        {Homans1958SocialExchange}
\bibfield{author}{\bibinfo{person}{George~Caspar. Homans}.} \bibinfo{year}{1958}\natexlab{}.
\newblock \showarticletitle{{Social Behavior as Exchange}}.
\newblock \bibinfo{journal}{\emph{https://doi.org/10.1086/222355}} \bibinfo{volume}{63}, \bibinfo{number}{6} (\bibinfo{date}{5} \bibinfo{year}{1958}), \bibinfo{pages}{597--606}.
\newblock
\showISSN{0002-9602}
\href{https://doi.org/10.1086/222355}{doi:\nolinkurl{10.1086/222355}}


\bibitem[Imel et~al\mbox{.}(2024)]%
        {Imel2024MentalLearning}
\bibfield{author}{\bibinfo{person}{Zac~E. Imel}, \bibinfo{person}{Michael~J. Tanana}, \bibinfo{person}{Christina~S. Soma}, \bibinfo{person}{Thomas~D. Hull}, \bibinfo{person}{Brian~T. Pace}, \bibinfo{person}{Sarah~C. Stanco}, \bibinfo{person}{Torrey~A. Creed}, \bibinfo{person}{Theresa~B. Moyers}, {and} \bibinfo{person}{David~C. Atkins}.} \bibinfo{year}{2024}\natexlab{}.
\newblock \showarticletitle{{Mental Health Counseling From Conversational Content With Transformer-Based Machine Learning}}.
\newblock \bibinfo{journal}{\emph{JAMA Network Open}} \bibinfo{volume}{7}, \bibinfo{number}{1} (\bibinfo{year}{2024}), \bibinfo{pages}{e2352590}.
\newblock
\showISSN{25743805}
\href{https://doi.org/10.1001/JAMANETWORKOPEN.2023.52590}{doi:\nolinkurl{10.1001/JAMANETWORKOPEN.2023.52590}}


\bibitem[Ireland et~al\mbox{.}(2011)]%
        {Ireland2011LanguageStability}
\bibfield{author}{\bibinfo{person}{Molly~E. Ireland}, \bibinfo{person}{Richard~B. Slatcher}, \bibinfo{person}{Paul~W. Eastwick}, \bibinfo{person}{Lauren~E. Scissors}, \bibinfo{person}{Eli~J. Finkel}, {and} \bibinfo{person}{James~W. Pennebaker}.} \bibinfo{year}{2011}\natexlab{}.
\newblock \showarticletitle{{Language Style Matching Predicts Relationship Initiation and Stability}}.
\newblock \bibinfo{journal}{\emph{Psychological Science}} \bibinfo{volume}{22}, \bibinfo{number}{1} (\bibinfo{date}{1} \bibinfo{year}{2011}), \bibinfo{pages}{39--44}.
\newblock
\showISSN{09567976}
\href{https://doi.org/10.1177/0956797610392928}{doi:\nolinkurl{10.1177/0956797610392928}}


\bibitem[Irfan and Skantze(2025)]%
        {irfan2025between}
\bibfield{author}{\bibinfo{person}{Bahar Irfan} {and} \bibinfo{person}{Gabriel Skantze}.} \bibinfo{year}{2025}\natexlab{}.
\newblock \showarticletitle{Between you and me: Ethics of self-disclosure in human-robot interaction}. In \bibinfo{booktitle}{\emph{Proceedings of the 2025 ACM/IEEE International Conference on Human-Robot Interaction}}. \bibinfo{pages}{1357--1362}.
\newblock


\bibitem[Jain et~al\mbox{.}(1999)]%
        {jain1999data}
\bibfield{author}{\bibinfo{person}{Anil~K Jain}, \bibinfo{person}{M~Narasimha Murty}, {and} \bibinfo{person}{Patrick~J Flynn}.} \bibinfo{year}{1999}\natexlab{}.
\newblock \showarticletitle{Data clustering: a review}.
\newblock \bibinfo{journal}{\emph{ACM computing surveys (CSUR)}} \bibinfo{volume}{31}, \bibinfo{number}{3} (\bibinfo{year}{1999}), \bibinfo{pages}{264--323}.
\newblock


\bibitem[Johnson et~al\mbox{.}(2024)]%
        {johnson2024detailed}
\bibfield{author}{\bibinfo{person}{S~Joshua Johnson}, \bibinfo{person}{M~Ramakrishna Murty}, {and} \bibinfo{person}{I Navakanth}.} \bibinfo{year}{2024}\natexlab{}.
\newblock \showarticletitle{A detailed review on word embedding techniques with emphasis on word2vec}.
\newblock \bibinfo{journal}{\emph{Multimedia Tools and Applications}} \bibinfo{volume}{83}, \bibinfo{number}{13} (\bibinfo{year}{2024}), \bibinfo{pages}{37979--38007}.
\newblock


\bibitem[Laban et~al\mbox{.}(2022)]%
        {Laban2022SocialTreatment}
\bibfield{author}{\bibinfo{person}{Guy Laban}, \bibinfo{person}{Ziv Ben-Zion}, {and} \bibinfo{person}{Emily~S. Cross}.} \bibinfo{year}{2022}\natexlab{}.
\newblock \showarticletitle{{Social Robots for Supporting Post-traumatic Stress Disorder Diagnosis and Treatment}}.
\newblock \bibinfo{journal}{\emph{Frontiers in psychiatry}}  \bibinfo{volume}{12} (\bibinfo{year}{2022}).
\newblock
\showISSN{1664-0640}
\href{https://doi.org/10.3389/FPSYT.2021.752874}{doi:\nolinkurl{10.3389/FPSYT.2021.752874}}


\bibitem[Laban et~al\mbox{.}(2025a)]%
        {Laban2025WhatTime}
\bibfield{author}{\bibinfo{person}{Guy Laban}, \bibinfo{person}{Sophie Chiang}, {and} \bibinfo{person}{Hatice Gunes}.} \bibinfo{year}{2025}\natexlab{a}.
\newblock \showarticletitle{{What People Share With a Robot When Feeling Lonely and Stressed and How It Helps Over Time}}. In \bibinfo{booktitle}{\emph{2025 34nd IEEE International Conference on Robot and Human Interactive Communication (RO-MAN)}}.
\newblock


\bibitem[Laban and Cross(2024)]%
        {Laban2024SharingFeel}
\bibfield{author}{\bibinfo{person}{Guy Laban} {and} \bibinfo{person}{Emily~S. Cross}.} \bibinfo{year}{2024}\natexlab{}.
\newblock \showarticletitle{{Sharing our Emotions with Robots: Why do we do it and how does it make us feel?}}
\newblock \bibinfo{journal}{\emph{IEEE Transactions on Affective Computing}} (\bibinfo{year}{2024}), \bibinfo{pages}{1--18}.
\newblock
\showISSN{1949-3045}
\href{https://doi.org/10.1109/TAFFC.2024.3470984}{doi:\nolinkurl{10.1109/TAFFC.2024.3470984}}


\bibitem[Laban et~al\mbox{.}(2021)]%
        {Laban2021}
\bibfield{author}{\bibinfo{person}{Guy Laban}, \bibinfo{person}{Jean-Noël George}, \bibinfo{person}{Val Morrison}, {and} \bibinfo{person}{Emily~S. Cross}.} \bibinfo{year}{2021}\natexlab{}.
\newblock \showarticletitle{{Tell me more! Assessing interactions with social robots from speech}}.
\newblock \bibinfo{journal}{\emph{Paladyn, Journal of Behavioral Robotics}} \bibinfo{volume}{12}, \bibinfo{number}{1} (\bibinfo{year}{2021}), \bibinfo{pages}{136--159}.
\newblock
\showISSN{20814836}
\href{https://doi.org/10.1515/pjbr-2021-0011}{doi:\nolinkurl{10.1515/pjbr-2021-0011}}


\bibitem[Laban et~al\mbox{.}(2023)]%
        {Laban2023OpeningBehavior}
\bibfield{author}{\bibinfo{person}{Guy Laban}, \bibinfo{person}{Arvid Kappas}, \bibinfo{person}{Val Morrison}, {and} \bibinfo{person}{Emily~S. Cross}.} \bibinfo{year}{2023}\natexlab{}.
\newblock \showarticletitle{{Opening Up to Social Robots: How Emotions Drive Self-Disclosure Behavior}}. In \bibinfo{booktitle}{\emph{2023 32nd IEEE International Conference on Robot and Human Interactive Communication (RO-MAN)}}. \bibinfo{publisher}{IEEE}, \bibinfo{address}{Busan, Republic of Korea}, \bibinfo{pages}{1697--1704}.
\newblock
\showISBNx{979-8-3503-3670-2}
\href{https://doi.org/10.1109/RO-MAN57019.2023.10309551}{doi:\nolinkurl{10.1109/RO-MAN57019.2023.10309551}}


\bibitem[Laban et~al\mbox{.}(2024a)]%
        {Laban2024BuildingTime}
\bibfield{author}{\bibinfo{person}{Guy Laban}, \bibinfo{person}{Arvid Kappas}, \bibinfo{person}{Val Morrison}, {and} \bibinfo{person}{Emily~S Cross}.} \bibinfo{year}{2024}\natexlab{a}.
\newblock \showarticletitle{{Building Long-Term Human–Robot Relationships: Examining Disclosure, Perception and Well-Being Across Time}}.
\newblock \bibinfo{journal}{\emph{International Journal of Social Robotics}} \bibinfo{volume}{16}, \bibinfo{number}{5} (\bibinfo{year}{2024}), \bibinfo{pages}{1--27}.
\newblock
\showISSN{1875-4805}
\href{https://doi.org/10.1007/s12369-023-01076-z}{doi:\nolinkurl{10.1007/s12369-023-01076-z}}


\bibitem[Laban et~al\mbox{.}(2024b)]%
        {laban_lexi_hai_24}
\bibfield{author}{\bibinfo{person}{Guy Laban}, \bibinfo{person}{Tomer Laban}, {and} \bibinfo{person}{Hatice Gunes}.} \bibinfo{year}{2024}\natexlab{b}.
\newblock \showarticletitle{{LEXI: Large Language Models Experimentation Interface}}. In \bibinfo{booktitle}{\emph{Proceedings of the 12th International Conference on Human-Agent Interaction}}. \bibinfo{publisher}{ACM}, \bibinfo{address}{New York, NY, USA}, \bibinfo{pages}{250--259}.
\newblock
\showISBNx{9798400711787}
\href{https://doi.org/10.1145/3687272.3688296}{doi:\nolinkurl{10.1145/3687272.3688296}}


\bibitem[Laban et~al\mbox{.}(2024c)]%
        {Laban2024SocialWell-Being}
\bibfield{author}{\bibinfo{person}{Guy Laban}, \bibinfo{person}{Val Morrison}, {and} \bibinfo{person}{Emily Cross}.} \bibinfo{year}{2024}\natexlab{c}.
\newblock \showarticletitle{{Social Robots for Health Psychology: A New Frontier for Improving Human Health and Well-Being}}.
\newblock \bibinfo{journal}{\emph{European Health Psychologist}} \bibinfo{volume}{23}, \bibinfo{number}{1} (\bibinfo{date}{2} \bibinfo{year}{2024}), \bibinfo{pages}{1095--1102}.
\newblock
\showISSN{2225-6962}
\urldef\tempurl%
\url{https://www.ehps.net/ehp/index.php/contents/article/view/3442}
\showURL{%
\tempurl}


\bibitem[Laban et~al\mbox{.}(2025b)]%
        {laban_ced_2023}
\bibfield{author}{\bibinfo{person}{Guy Laban}, \bibinfo{person}{Val Morrison}, \bibinfo{person}{Arvid Kappas}, {and} \bibinfo{person}{Emily~S. Cross}.} \bibinfo{year}{2025}\natexlab{b}.
\newblock \showarticletitle{{Coping with Emotional Distress via Self-Disclosure to Robots: An Intervention with Caregivers}}.
\newblock \bibinfo{journal}{\emph{International Journal of Social Robotics}} (\bibinfo{year}{2025}).
\newblock
\href{https://doi.org/10.1007/s12369-024-01207-0}{doi:\nolinkurl{10.1007/s12369-024-01207-0}}


\bibitem[Laban et~al\mbox{.}(2025c)]%
        {Laban2025CriticalWellbeing}
\bibfield{author}{\bibinfo{person}{Guy Laban}, \bibinfo{person}{Micol Spitale}, \bibinfo{person}{Minja Axelsson}, \bibinfo{person}{Nida~Itrat Abbasi}, {and} \bibinfo{person}{Hatice Gunes}.} \bibinfo{year}{2025}\natexlab{c}.
\newblock \showarticletitle{Critical Insights about Robots for Mental Wellbeing}.
\newblock  (\bibinfo{date}{6} \bibinfo{year}{2025}).
\newblock
\urldef\tempurl%
\url{https://arxiv.org/pdf/2506.13739}
\showURL{%
\tempurl}


\bibitem[Laban et~al\mbox{.}(2025d)]%
        {laban2025robot}
\bibfield{author}{\bibinfo{person}{Guy Laban}, \bibinfo{person}{Julie Wang}, {and} \bibinfo{person}{Hatice Gunes}.} \bibinfo{year}{2025}\natexlab{d}.
\newblock \showarticletitle{A Robot-Led Intervention for Emotion Regulation: From Expression to Reappraisal}.
\newblock \bibinfo{journal}{\emph{arXiv preprint arXiv:2503.18243}} (\bibinfo{year}{2025}).
\newblock


\bibitem[Lawler(2001)]%
        {RefWorks:262}
\bibfield{author}{\bibinfo{person}{Edward~J Lawler}.} \bibinfo{year}{2001}\natexlab{}.
\newblock \showarticletitle{{An Affect Theory of Social Exchange}}.
\newblock \bibinfo{journal}{\emph{Amer. J. Sociology}} \bibinfo{volume}{107}, \bibinfo{number}{2} (\bibinfo{year}{2001}), \bibinfo{pages}{321--352}.
\newblock
\showISBNx{0002-9602}
\href{https://doi.org/10.1086/324071}{doi:\nolinkurl{10.1086/324071}}


\bibitem[Likas et~al\mbox{.}(2003)]%
        {likas2003global}
\bibfield{author}{\bibinfo{person}{Aristidis Likas}, \bibinfo{person}{Nikos Vlassis}, {and} \bibinfo{person}{Jakob~J Verbeek}.} \bibinfo{year}{2003}\natexlab{}.
\newblock \showarticletitle{The global k-means clustering algorithm}.
\newblock \bibinfo{journal}{\emph{Pattern recognition}} \bibinfo{volume}{36}, \bibinfo{number}{2} (\bibinfo{year}{2003}), \bibinfo{pages}{451--461}.
\newblock


\bibitem[Liu and Deng(2020)]%
        {liu2020determine}
\bibfield{author}{\bibinfo{person}{Fan Liu} {and} \bibinfo{person}{Yong Deng}.} \bibinfo{year}{2020}\natexlab{}.
\newblock \showarticletitle{Determine the number of unknown targets in open world based on elbow method}.
\newblock \bibinfo{journal}{\emph{IEEE Transactions on Fuzzy Systems}} \bibinfo{volume}{29}, \bibinfo{number}{5} (\bibinfo{year}{2020}), \bibinfo{pages}{986--995}.
\newblock


\bibitem[Madani et~al\mbox{.}(2024)]%
        {Madani2024SteeringConversations}
\bibfield{author}{\bibinfo{person}{Navid Madani}, \bibinfo{person}{Sougata Saha}, {and} \bibinfo{person}{Rohini Srihari}.} \bibinfo{year}{2024}\natexlab{}.
\newblock \showarticletitle{{Steering Conversational Large Language Models for Long Emotional Support Conversations}}.
\newblock  (\bibinfo{date}{2} \bibinfo{year}{2024}).
\newblock
\urldef\tempurl%
\url{https://arxiv.org/abs/2402.10453v2}
\showURL{%
\tempurl}


\bibitem[Mehra et~al\mbox{.}(2025)]%
        {Mehra2025BeyondValues}
\bibfield{author}{\bibinfo{person}{Vaibhav Mehra}, \bibinfo{person}{Guy Laban}, {and} \bibinfo{person}{Hatice Gunes}.} \bibinfo{year}{2025}\natexlab{}.
\newblock \showarticletitle{{How Large Language Models Classify and Semantically Explain Facial Expressions from Valence-Arousal Values}}. In \bibinfo{booktitle}{\emph{Proceedings of the 7th ACM Conference on Conversational User Interfaces (CUI ’25)}}. \bibinfo{publisher}{Association for Computing Machinery}, \bibinfo{address}{New York, NY, USA}.
\newblock
\href{https://doi.org/10.1145/3719160.3737618}{doi:\nolinkurl{10.1145/3719160.3737618}}


\bibitem[Mikolov et~al\mbox{.}(2013)]%
        {Mikolov2013DistributedCompositionality}
\bibfield{author}{\bibinfo{person}{Tomas Mikolov}, \bibinfo{person}{Ilya Sutskever}, \bibinfo{person}{Kai Chen}, \bibinfo{person}{Greg~S. Corrado}, {and} \bibinfo{person}{Jeff Dean}.} \bibinfo{year}{2013}\natexlab{}.
\newblock \showarticletitle{{Distributed Representations of Words and Phrases and their Compositionality}}.
\newblock \bibinfo{journal}{\emph{NIPS}}  \bibinfo{volume}{26} (\bibinfo{year}{2013}).
\newblock


\bibitem[Miner et~al\mbox{.}(2022)]%
        {Miner2022AConsistency}
\bibfield{author}{\bibinfo{person}{Adam~S. Miner}, \bibinfo{person}{Scott~L. Fleming}, \bibinfo{person}{Albert Haque}, \bibinfo{person}{Jason~A. Fries}, \bibinfo{person}{Tim Althoff}, \bibinfo{person}{Denise~E. Wilfley}, \bibinfo{person}{W.~Stewart Agras}, \bibinfo{person}{Arnold Milstein}, \bibinfo{person}{Jeff Hancock}, \bibinfo{person}{Steven~M. Asch}, \bibinfo{person}{Shannon~Wiltsey Stirman}, \bibinfo{person}{Bruce~A. Arnow}, {and} \bibinfo{person}{Nigam~H. Shah}.} \bibinfo{year}{2022}\natexlab{}.
\newblock \showarticletitle{{A computational approach to measure the linguistic characteristics of psychotherapy timing, responsiveness, and consistency}}.
\newblock \bibinfo{journal}{\emph{npj Mental Health Research 2022 1:1}} \bibinfo{volume}{1}, \bibinfo{number}{1} (\bibinfo{date}{12} \bibinfo{year}{2022}), \bibinfo{pages}{1--12}.
\newblock
\showISSN{2731-4251}
\href{https://doi.org/10.1038/s44184-022-00020-9}{doi:\nolinkurl{10.1038/s44184-022-00020-9}}


\bibitem[Niederhoffer and Pennebaker(2002)]%
        {Niederhoffer2002LinguisticInteraction}
\bibfield{author}{\bibinfo{person}{Kate~G. Niederhoffer} {and} \bibinfo{person}{James~W. Pennebaker}.} \bibinfo{year}{2002}\natexlab{}.
\newblock \showarticletitle{{Linguistic Style Matching in Social Interaction}}.
\newblock \bibinfo{journal}{\emph{Journal of Language and Social Psychology}} \bibinfo{volume}{21}, \bibinfo{number}{4} (\bibinfo{date}{12} \bibinfo{year}{2002}), \bibinfo{pages}{337--360}.
\newblock
\showISSN{0261927X}
\href{https://doi.org/10.1177/026192702237953}{doi:\nolinkurl{10.1177/026192702237953}}


\bibitem[Nomura et~al\mbox{.}(2020)]%
        {Nomura2020}
\bibfield{author}{\bibinfo{person}{Tatsuya Nomura}, \bibinfo{person}{Takayuki Kanda}, \bibinfo{person}{Tomohiro Suzuki}, {and} \bibinfo{person}{Sachie Yamada}.} \bibinfo{year}{2020}\natexlab{}.
\newblock \showarticletitle{{Do people with social anxiety feel anxious about interacting with a robot?}}
\newblock \bibinfo{journal}{\emph{AI {\&} SOCIETY}} \bibinfo{volume}{35}, \bibinfo{number}{2} (\bibinfo{year}{2020}), \bibinfo{pages}{381--390}.
\newblock
\showISSN{1435-5655}
\href{https://doi.org/10.1007/s00146-019-00889-9}{doi:\nolinkurl{10.1007/s00146-019-00889-9}}


\bibitem[Ong et~al\mbox{.}(2023)]%
        {Ong2023PowerResearch}
\bibfield{author}{\bibinfo{person}{Ben Ong}, \bibinfo{person}{Eleftheria Tseliou}, \bibinfo{person}{Tom Strong}, {and} \bibinfo{person}{Niels Buus}.} \bibinfo{year}{2023}\natexlab{}.
\newblock \showarticletitle{{Power and dialogue: A review of discursive research}}.
\newblock \bibinfo{journal}{\emph{Family Process}} \bibinfo{volume}{62}, \bibinfo{number}{4} (\bibinfo{date}{12} \bibinfo{year}{2023}), \bibinfo{pages}{1391--1407}.
\newblock
\showISSN{1545-5300}
\href{https://doi.org/10.1111/FAMP.12881}{doi:\nolinkurl{10.1111/FAMP.12881}}


\bibitem[OpenAI(2024)]%
        {openai2024chatgpt}
\bibfield{author}{\bibinfo{person}{OpenAI}.} \bibinfo{year}{2024}\natexlab{}.
\newblock \bibinfo{title}{ChatGPT 40-mini}.
\newblock \bibinfo{howpublished}{\url{https://openai.com}}.
\newblock
\newblock
\shownote{Large language model. Accessed November 13, 2024.}.


\bibitem[Pedregosa et~al\mbox{.}(2011)]%
        {PedregosaFABIANPEDREGOSA2011Scikit-learn:Python}
\bibfield{author}{\bibinfo{person}{Fabian Pedregosa}, \bibinfo{person}{Vincent Michel}, \bibinfo{person}{Olivier Grisel}, \bibinfo{person}{Mathieu Blondel}, \bibinfo{person}{Peter Prettenhofer}, \bibinfo{person}{Ron Weiss}, \bibinfo{person}{Jake Vanderplas}, \bibinfo{person}{David Cournapeau}, \bibinfo{person}{Gaël Varoquaux}, \bibinfo{person}{Alexandre Gramfort}, \bibinfo{person}{Bertrand Thirion}, \bibinfo{person}{Vincent Dubourg}, \bibinfo{person}{Alexandre Passos}, \bibinfo{person}{Matthieu Brucher}, \bibinfo{person}{Matthieu {\'{E}}douardand}, {and} \bibinfo{person}{Édouard Duchesnay}.} \bibinfo{year}{2011}\natexlab{}.
\newblock \showarticletitle{{Scikit-learn: Machine Learning in Python}}.
\newblock \bibinfo{journal}{\emph{The Journal of Machine Learning Research}}  \bibinfo{volume}{12} (\bibinfo{date}{11} \bibinfo{year}{2011}), \bibinfo{pages}{2825--2830}.
\newblock
\href{https://doi.org/10.5555/1953048.2078195}{doi:\nolinkurl{10.5555/1953048.2078195}}


\bibitem[Perez-Osorio et~al\mbox{.}(2015)]%
        {Perez-Osorio2015GazeGoals}
\bibfield{author}{\bibinfo{person}{Jairo Perez-Osorio}, \bibinfo{person}{Hermann~J. M{\"{u}}ller}, \bibinfo{person}{Eva Wiese}, {and} \bibinfo{person}{Agnieszka Wykowska}.} \bibinfo{year}{2015}\natexlab{}.
\newblock \showarticletitle{{Gaze Following Is Modulated by Expectations Regarding Others’ Action Goals}}.
\newblock \bibinfo{journal}{\emph{PLOS ONE}} \bibinfo{volume}{10}, \bibinfo{number}{11} (\bibinfo{date}{11} \bibinfo{year}{2015}), \bibinfo{pages}{e0143614}.
\newblock
\showISSN{1932-6203}
\href{https://doi.org/10.1371/JOURNAL.PONE.0143614}{doi:\nolinkurl{10.1371/JOURNAL.PONE.0143614}}


\bibitem[Perez-Osorio et~al\mbox{.}(2017)]%
        {Perez-Osorio2017ExpectationsEffect}
\bibfield{author}{\bibinfo{person}{Jairo Perez-Osorio}, \bibinfo{person}{Hermann~J. M{\"{u}}ller}, {and} \bibinfo{person}{Agnieszka Wykowska}.} \bibinfo{year}{2017}\natexlab{}.
\newblock \showarticletitle{{Expectations regarding action sequences modulate electrophysiological correlates of the gaze-cueing effect}}.
\newblock \bibinfo{journal}{\emph{Psychophysiology}} \bibinfo{volume}{54}, \bibinfo{number}{7} (\bibinfo{date}{7} \bibinfo{year}{2017}), \bibinfo{pages}{942--954}.
\newblock
\showISSN{1469-8986}
\href{https://doi.org/10.1111/PSYP.12854}{doi:\nolinkurl{10.1111/PSYP.12854}}


\bibitem[Pu et~al\mbox{.}(2020)]%
        {pu2020hybrid}
\bibfield{author}{\bibinfo{person}{Guo Pu}, \bibinfo{person}{Lijuan Wang}, \bibinfo{person}{Jun Shen}, {and} \bibinfo{person}{Fang Dong}.} \bibinfo{year}{2020}\natexlab{}.
\newblock \showarticletitle{A hybrid unsupervised clustering-based anomaly detection method}.
\newblock \bibinfo{journal}{\emph{Tsinghua Science and Technology}} \bibinfo{volume}{26}, \bibinfo{number}{2} (\bibinfo{year}{2020}), \bibinfo{pages}{146--153}.
\newblock


\bibitem[Ragelien{\.{e}}(2016)]%
        {Rageliene2016LinksReview}
\bibfield{author}{\bibinfo{person}{Tija Ragelien{\.{e}}}.} \bibinfo{year}{2016}\natexlab{}.
\newblock \showarticletitle{{Links of Adolescents Identity Development and Relationship with Peers: A Systematic Literature Review}}.
\newblock \bibinfo{journal}{\emph{Journal of the Canadian Academy of Child and Adolescent Psychiatry}} \bibinfo{volume}{25}, \bibinfo{number}{2} (\bibinfo{date}{3} \bibinfo{year}{2016}), \bibinfo{pages}{97}.
\newblock
\showISSN{17198429}
\urldef\tempurl%
\url{https://pmc.ncbi.nlm.nih.gov/articles/PMC4879949/}
\showURL{%
\tempurl}


\bibitem[Reece et~al\mbox{.}(2023)]%
        {Reece2023TheConversation}
\bibfield{author}{\bibinfo{person}{Andrew Reece}, \bibinfo{person}{Gus Cooney}, \bibinfo{person}{Peter Bull}, \bibinfo{person}{Christine Chung}, \bibinfo{person}{Bryn Dawson}, \bibinfo{person}{Casey Fitzpatrick}, \bibinfo{person}{Tamara Glazer}, \bibinfo{person}{Dean Knox}, \bibinfo{person}{Alex Liebscher}, {and} \bibinfo{person}{Sebastian Marin}.} \bibinfo{year}{2023}\natexlab{}.
\newblock \showarticletitle{{The CANDOR corpus: Insights from a large multimodal dataset of naturalistic conversation}}.
\newblock \bibinfo{journal}{\emph{Science Advances}} \bibinfo{volume}{9}, \bibinfo{number}{13} (\bibinfo{date}{3} \bibinfo{year}{2023}).
\newblock
\showISSN{23752548}
\href{https://doi.org/doi.org/10.1126/sciadv.adf3197}{doi:\nolinkurl{doi.org/10.1126/sciadv.adf3197}}


\bibitem[Rehm et~al\mbox{.}(2007)]%
        {rehm2007novel}
\bibfield{author}{\bibinfo{person}{Frank Rehm}, \bibinfo{person}{Frank Klawonn}, {and} \bibinfo{person}{Rudolf Kruse}.} \bibinfo{year}{2007}\natexlab{}.
\newblock \showarticletitle{A novel approach to noise clustering for outlier detection}.
\newblock \bibinfo{journal}{\emph{Soft Computing}}  \bibinfo{volume}{11} (\bibinfo{year}{2007}), \bibinfo{pages}{489--494}.
\newblock


\bibitem[Reimers and Gurevych(2019)]%
        {Reimers2019Sentence-BERT:BERT-Networks}
\bibfield{author}{\bibinfo{person}{Nils Reimers} {and} \bibinfo{person}{Iryna Gurevych}.} \bibinfo{year}{2019}\natexlab{}.
\newblock \showarticletitle{{Sentence-BERT: Sentence Embeddings using Siamese BERT-Networks}}.
\newblock \bibinfo{journal}{\emph{Proceedings of EMNLP-IJCNLP 2019}} (\bibinfo{year}{2019}), \bibinfo{pages}{3982--3992}.
\newblock
\showISBNx{9781950737901}
\href{https://doi.org/10.18653/V1/D19-1410}{doi:\nolinkurl{10.18653/V1/D19-1410}}


\bibitem[Riddoch and Cross(2023)]%
        {Riddoch2023InvestigatingRobot}
\bibfield{author}{\bibinfo{person}{Katie~A. Riddoch} {and} \bibinfo{person}{Emily~S. Cross}.} \bibinfo{year}{2023}\natexlab{}.
\newblock \showarticletitle{{Investigating the effect of cardio-visual synchrony on prosocial behavior towards a social robot}}.
\newblock \bibinfo{journal}{\emph{Open Research Europe}}  \bibinfo{volume}{3} (\bibinfo{date}{2} \bibinfo{year}{2023}), \bibinfo{pages}{37}.
\newblock
\href{https://doi.org/10.12688/OPENRESEUROPE.15003.1}{doi:\nolinkurl{10.12688/OPENRESEUROPE.15003.1}}


\bibitem[Robinson et~al\mbox{.}(2019)]%
        {RefWorks:404}
\bibfield{author}{\bibinfo{person}{Nicole~Lee Robinson}, \bibinfo{person}{Timothy~Vaughan Cottier}, {and} \bibinfo{person}{David~John Kavanagh}.} \bibinfo{year}{2019}\natexlab{}.
\newblock \showarticletitle{{Psychosocial Health Interventions by Social Robots: Systematic Review of Randomized Controlled Trials}}.
\newblock \bibinfo{journal}{\emph{J Med Internet Res}} \bibinfo{volume}{21}, \bibinfo{number}{5} (\bibinfo{year}{2019}), \bibinfo{pages}{1--20}.
\newblock
\showISBNx{1438-8871}
\href{https://doi.org/10.2196/13203}{doi:\nolinkurl{10.2196/13203}}


\bibitem[Rogers(1951)]%
        {Rogers1951Client-centeredTheory.}
\bibfield{author}{\bibinfo{person}{Carl~R Rogers}.} \bibinfo{year}{1951}\natexlab{}.
\newblock \bibinfo{booktitle}{\emph{{Client-centered therapy; its current practice, implications, and theory.}}}
\newblock \bibinfo{publisher}{Houghton Mifflin}, \bibinfo{address}{Oxford, England}. 560, xii, 560--xii pages.
\newblock


\bibitem[Scoglio et~al\mbox{.}(2019)]%
        {RefWorks:424}
\bibfield{author}{\bibinfo{person}{Arielle A~J Scoglio}, \bibinfo{person}{Erin~D Reilly}, \bibinfo{person}{Jay~A Gorman}, {and} \bibinfo{person}{Charles~E Drebing}.} \bibinfo{year}{2019}\natexlab{}.
\newblock \showarticletitle{{Use of Social Robots in Mental Health and Well-Being Research: Systematic Review}}.
\newblock \bibinfo{journal}{\emph{J Med Internet Res}} \bibinfo{volume}{21}, \bibinfo{number}{7} (\bibinfo{year}{2019}), \bibinfo{pages}{e13322}.
\newblock
\showISBNx{1438-8871}
\href{https://doi.org/10.2196/13322}{doi:\nolinkurl{10.2196/13322}}


\bibitem[Scotti et~al\mbox{.}(2023)]%
        {Scotti2023AChatbots}
\bibfield{author}{\bibinfo{person}{Vincenzo Scotti}, \bibinfo{person}{Licia Sbattella}, {and} \bibinfo{person}{Roberto Tedesco}.} \bibinfo{year}{2023}\natexlab{}.
\newblock \showarticletitle{{A Primer on Seq2Seq Models for Generative Chatbots}}.
\newblock \bibinfo{journal}{\emph{Comput. Surveys}} \bibinfo{volume}{56}, \bibinfo{number}{3} (\bibinfo{date}{3} \bibinfo{year}{2023}).
\newblock
\showISSN{15577341}
\href{https://doi.org/10.1145/3604281/ASSET/C6FFEB58-CD32-45F1-B2AE-4655B3610219/ASSETS/GRAPHIC/CSUR-2022-0383-F19.JPG}{doi:\nolinkurl{10.1145/3604281/ASSET/C6FFEB58-CD32-45F1-B2AE-4655B3610219/ASSETS/GRAPHIC/CSUR-2022-0383-F19.JPG}}


\bibitem[Seligman(2018)]%
        {Seligman2018PERMAWell-being}
\bibfield{author}{\bibinfo{person}{Martin Elias~Peter Seligman}.} \bibinfo{year}{2018}\natexlab{}.
\newblock \showarticletitle{{PERMA and the building blocks of well-being}}.
\newblock \bibinfo{journal}{\emph{The Journal of Positive Psychology}} \bibinfo{volume}{13}, \bibinfo{number}{4} (\bibinfo{date}{7} \bibinfo{year}{2018}), \bibinfo{pages}{333--335}.
\newblock
\showISSN{17439779}
\href{https://doi.org/10.1080/17439760.2018.1437466}{doi:\nolinkurl{10.1080/17439760.2018.1437466}}


\bibitem[Selva~Birunda and Kanniga~Devi(2021)]%
        {selva2021review}
\bibfield{author}{\bibinfo{person}{S Selva~Birunda} {and} \bibinfo{person}{R Kanniga~Devi}.} \bibinfo{year}{2021}\natexlab{}.
\newblock \showarticletitle{A review on word embedding techniques for text classification}.
\newblock \bibinfo{journal}{\emph{Innovative Data Communication Technologies and Application: Proceedings of ICIDCA 2020}} (\bibinfo{year}{2021}), \bibinfo{pages}{267--281}.
\newblock


\bibitem[Sharma et~al\mbox{.}(2020)]%
        {Sharma2020ASupport}
\bibfield{author}{\bibinfo{person}{Ashish Sharma}, \bibinfo{person}{Adam~S Miner}, \bibinfo{person}{David~C Atkins}, \bibinfo{person}{Tim Althoff}, {and} \bibinfo{person}{Paul~G Allen}.} \bibinfo{year}{2020}\natexlab{}.
\newblock \showarticletitle{{A Computational Approach to Understanding Empathy Expressed in Text-Based Mental Health Support}}.
\newblock   \bibinfo{volume}{5263} (\bibinfo{date}{11} \bibinfo{year}{2020}), \bibinfo{pages}{5263--5276}.
\newblock
\href{https://doi.org/10.18653/V1/2020.EMNLP-MAIN.425}{doi:\nolinkurl{10.18653/V1/2020.EMNLP-MAIN.425}}


\bibitem[Skantze(2021)]%
        {Skantze2021Turn-takingReview}
\bibfield{author}{\bibinfo{person}{Gabriel Skantze}.} \bibinfo{year}{2021}\natexlab{}.
\newblock \showarticletitle{{Turn-taking in Conversational Systems and Human-Robot Interaction: A Review}}.
\newblock \bibinfo{journal}{\emph{Computer Speech {\&} Language}}  \bibinfo{volume}{67} (\bibinfo{date}{5} \bibinfo{year}{2021}), \bibinfo{pages}{101178}.
\newblock
\showISSN{0885-2308}
\href{https://doi.org/10.1016/J.CSL.2020.101178}{doi:\nolinkurl{10.1016/J.CSL.2020.101178}}


\bibitem[Skantze and Irfan(2025)]%
        {10.5555/3721488.3721593}
\bibfield{author}{\bibinfo{person}{Gabriel Skantze} {and} \bibinfo{person}{Bahar Irfan}.} \bibinfo{year}{2025}\natexlab{}.
\newblock \showarticletitle{Applying General Turn-taking Models to Conversational Human-Robot Interaction}. In \bibinfo{booktitle}{\emph{Proceedings of the 2025 ACM/IEEE International Conference on Human-Robot Interaction}} (Melbourne, Australia) \emph{(\bibinfo{series}{HRI '25})}. \bibinfo{publisher}{IEEE Press}, \bibinfo{pages}{859–868}.
\newblock


\bibitem[Soman et~al\mbox{.}(2025)]%
        {Soman2025HumanGeneration}
\bibfield{author}{\bibinfo{person}{Gayathri Soman}, \bibinfo{person}{M.~V. Judy}, {and} \bibinfo{person}{Aadhil~Muhammad Abou}.} \bibinfo{year}{2025}\natexlab{}.
\newblock \showarticletitle{{Human guided empathetic AI agent for mental health support leveraging reinforcement learning-enhanced retrieval-augmented generation}}.
\newblock \bibinfo{journal}{\emph{Cognitive Systems Research}}  \bibinfo{volume}{90} (\bibinfo{date}{4} \bibinfo{year}{2025}), \bibinfo{pages}{101337}.
\newblock
\showISSN{1389-0417}
\href{https://doi.org/10.1016/J.COGSYS.2025.101337}{doi:\nolinkurl{10.1016/J.COGSYS.2025.101337}}


\bibitem[Spitale et~al\mbox{.}(2024)]%
        {Spitale2024AppropriatenessEvaluation}
\bibfield{author}{\bibinfo{person}{Micol Spitale}, \bibinfo{person}{Minja Axelsson}, {and} \bibinfo{person}{Hatice Gunes}.} \bibinfo{year}{2024}\natexlab{}.
\newblock \showarticletitle{{Appropriateness of LLM-equipped Robotic Well-being Coach Language in the Workplace: A Qualitative Evaluation}}.
\newblock  (\bibinfo{date}{1} \bibinfo{year}{2024}).
\newblock
\urldef\tempurl%
\url{https://arxiv.org/abs/2401.14935v1}
\showURL{%
\tempurl}


\bibitem[Spitale et~al\mbox{.}(2025)]%
        {Spitale2024PastWell-being}
\bibfield{author}{\bibinfo{person}{Micol Spitale}, \bibinfo{person}{Minja Axelsson}, \bibinfo{person}{Sooyeon Jeong}, \bibinfo{person}{Paige Tuttösí}, \bibinfo{person}{Caitlin~A. Stamatis}, \bibinfo{person}{Guy Laban}, \bibinfo{person}{Angelica Lim}, {and} \bibinfo{person}{Hatice Gunes}.} \bibinfo{year}{2025}\natexlab{}.
\newblock \showarticletitle{Past, Present, and Future: A Survey of The Evolution of Affective Robotics For Well-being}.
\newblock \bibinfo{journal}{\emph{IEEE Transactions on Affective Computing}} (\bibinfo{year}{2025}), \bibinfo{pages}{1--17}.
\newblock
\showISSN{1949-3045}
\href{https://doi.org/10.1109/TAFFC.2025.3567740}{doi:\nolinkurl{10.1109/TAFFC.2025.3567740}}


\bibitem[{Spitale Micol} et~al\mbox{.}(2025)]%
        {SpitaleMicol2025VITA:Coaching}
\bibfield{author}{\bibinfo{person}{{Spitale Micol}}, \bibinfo{person}{{Axelsson Minja}}, {and} \bibinfo{person}{{Gunes Hatice}}.} \bibinfo{year}{2025}\natexlab{}.
\newblock \showarticletitle{{VITA: A Multi-Modal LLM-Based System for Longitudinal, Autonomous and Adaptive Robotic Mental Well-Being Coaching}}.
\newblock \bibinfo{journal}{\emph{ACM Transactions on Human-Robot Interaction}} \bibinfo{volume}{14}, \bibinfo{number}{2} (\bibinfo{date}{3} \bibinfo{year}{2025}), \bibinfo{pages}{1--28}.
\newblock
\showISBNx{10.1145/3712265}
\showISSN{2573-9522}
\href{https://doi.org/10.1145/3712265}{doi:\nolinkurl{10.1145/3712265}}


\bibitem[Tak and Gratch(2023)]%
        {Tak2023IsEmotion}
\bibfield{author}{\bibinfo{person}{Ala~N. Tak} {and} \bibinfo{person}{Jonathan Gratch}.} \bibinfo{year}{2023}\natexlab{}.
\newblock \showarticletitle{{Is GPT a Computational Model of Emotion?}}
\newblock \bibinfo{journal}{\emph{2023 11th International Conference on Affective Computing and Intelligent Interaction (ACII)}} (\bibinfo{date}{9} \bibinfo{year}{2023}), \bibinfo{pages}{1--8}.
\newblock
\showISBNx{979-8-3503-2743-4}
\href{https://doi.org/10.1109/ACII59096.2023.10388119}{doi:\nolinkurl{10.1109/ACII59096.2023.10388119}}


\bibitem[Tak and Gratch(2024)]%
        {Tak2024GPT-4Perspective}
\bibfield{author}{\bibinfo{person}{Ala~N. Tak} {and} \bibinfo{person}{Jonathan Gratch}.} \bibinfo{year}{2024}\natexlab{}.
\newblock \showarticletitle{{GPT-4 Emulates Average-Human Emotional Cognition from a Third-Person Perspective}}.
\newblock  (\bibinfo{date}{8} \bibinfo{year}{2024}).
\newblock
\urldef\tempurl%
\url{https://arxiv.org/abs/2408.13718v1}
\showURL{%
\tempurl}


\bibitem[Templeton et~al\mbox{.}(2022)]%
        {Templeton2022FastConversation}
\bibfield{author}{\bibinfo{person}{Emma~M. Templeton}, \bibinfo{person}{Luke~J. Chang}, \bibinfo{person}{Elizabeth~A. Reynolds}, \bibinfo{person}{Marie~D.Cone LeBeaumont}, {and} \bibinfo{person}{Thalia Wheatley}.} \bibinfo{year}{2022}\natexlab{}.
\newblock \showarticletitle{{Fast response times signal social connection in conversation}}.
\newblock \bibinfo{journal}{\emph{Proceedings of the National Academy of Sciences of the United States of America}} \bibinfo{volume}{119}, \bibinfo{number}{4} (\bibinfo{date}{1} \bibinfo{year}{2022}), \bibinfo{pages}{e2116915119}.
\newblock
\showISSN{10916490}
\href{https://doi.org/10.1073/PNAS.2116915119/SUPPL{\_}FILE/PNAS.2116915119.SAPP.PDF}{doi:\nolinkurl{10.1073/PNAS.2116915119/SUPPL{\_}FILE/PNAS.2116915119.SAPP.PDF}}


\bibitem[Tickle-Degnen and Rosenthal(1990)]%
        {Tickle-Degnen1990}
\bibfield{author}{\bibinfo{person}{Linda Tickle-Degnen} {and} \bibinfo{person}{Robert Rosenthal}.} \bibinfo{year}{1990}\natexlab{}.
\newblock \showarticletitle{{The Nature of Rapport and Its Nonverbal Correlates}}.
\newblock \bibinfo{journal}{\emph{Psychological Inquiry}} \bibinfo{volume}{1}, \bibinfo{number}{4} (\bibinfo{date}{1} \bibinfo{year}{1990}), \bibinfo{pages}{285--293}.
\newblock
\showISSN{15327965}
\href{https://doi.org/10.1207/S15327965PLI0104{\_}1}{doi:\nolinkurl{10.1207/S15327965PLI0104{\_}1}}


\bibitem[Vaswani et~al\mbox{.}(2017)]%
        {10.5555/3295222.3295349}
\bibfield{author}{\bibinfo{person}{Ashish Vaswani}, \bibinfo{person}{Noam Shazeer}, \bibinfo{person}{Niki Parmar}, \bibinfo{person}{Jakob Uszkoreit}, \bibinfo{person}{Llion Jones}, \bibinfo{person}{Aidan~N. Gomez}, \bibinfo{person}{\L{}ukasz Kaiser}, {and} \bibinfo{person}{Illia Polosukhin}.} \bibinfo{year}{2017}\natexlab{}.
\newblock \showarticletitle{Attention is all you need}. In \bibinfo{booktitle}{\emph{NIPS}} (Long Beach, California, USA). \bibinfo{pages}{6000–6010}.
\newblock
\showISBNx{9781510860964}


\bibitem[Wang et~al\mbox{.}(2020)]%
        {10.5555/3495724.3496209}
\bibfield{author}{\bibinfo{person}{Wenhui Wang}, \bibinfo{person}{Furu Wei}, \bibinfo{person}{Li Dong}, \bibinfo{person}{Hangbo Bao}, \bibinfo{person}{Nan Yang}, {and} \bibinfo{person}{Ming Zhou}.} \bibinfo{year}{2020}\natexlab{}.
\newblock \showarticletitle{MINILM: deep self-attention distillation for task-agnostic compression of pre-trained transformers}. In \bibinfo{booktitle}{\emph{NIPS}} (Vancouver, BC, Canada). Article \bibinfo{articleno}{485}, \bibinfo{numpages}{13}~pages.
\newblock
\showISBNx{9781713829546}


\bibitem[Yang et~al\mbox{.}(2019)]%
        {10.1145/3290605.3300261}
\bibfield{author}{\bibinfo{person}{Diyi Yang}, \bibinfo{person}{Zheng Yao}, \bibinfo{person}{Joseph Seering}, {and} \bibinfo{person}{Robert Kraut}.} \bibinfo{year}{2019}\natexlab{}.
\newblock \showarticletitle{The Channel Matters: Self-disclosure, Reciprocity and Social Support in Online Cancer Support Groups}. In \bibinfo{booktitle}{\emph{Proceedings of the 2019 CHI Conference on Human Factors in Computing Systems}} (Glasgow, Scotland Uk) \emph{(\bibinfo{series}{CHI '19})}. \bibinfo{publisher}{Association for Computing Machinery}, \bibinfo{address}{New York, NY, USA}, \bibinfo{pages}{1–15}.
\newblock
\showISBNx{9781450359702}
\href{https://doi.org/10.1145/3290605.3300261}{doi:\nolinkurl{10.1145/3290605.3300261}}


\bibitem[Yin et~al\mbox{.}(2016)]%
        {Yin2016PrayForDad:Information}
\bibfield{author}{\bibinfo{person}{Zhijun Yin}, \bibinfo{person}{You Chen}, \bibinfo{person}{Daniel Fabbri}, \bibinfo{person}{Jimeng Sun}, {and} \bibinfo{person}{Bradley Malin}.} \bibinfo{year}{2016}\natexlab{}.
\newblock \showarticletitle{{{\#}PrayForDad: Learning the Semantics Behind Why Social Media Users Disclose Health Information}}.
\newblock \bibinfo{journal}{\emph{Proceedings of the International AAAI Conference on Weblogs and Social Media.}}  \bibinfo{volume}{2016} (\bibinfo{year}{2016}), \bibinfo{pages}{456}.
\newblock
\showISBNx{9781577357582}
\showISSN{2162-3449}
\href{https://doi.org/10.1609/icwsm.v10i1.14735}{doi:\nolinkurl{10.1609/icwsm.v10i1.14735}}


\end{thebibliography}

\appendix

\begin{table*}[ht]
\centering
\caption{LLM generated Cluster Labels and Descriptions for human disclosures to human (therapist)}
\label{tab:h2h_clusters}
\scalebox{0.85}{
\begin{tabular}{|p{1.5cm}|p{15cm}|}
\hline

\textbf{Cluster Label} & \textbf{Cluster Description} \\ \hline
Family Conflict and Emotional Strain &
This cluster reveals deep-seated family conflicts characterized by controlling behaviors, communication breakdowns, and emotional distress. Respondents express frustration and helplessness regarding their relationships with parental figures and their children. The recurring theme of control and emotional manipulation is evident, particularly in the strained dynamics between adults and their mothers or partners. Concern is raised over the inability to communicate effectively, leading to feelings of resentment and blame, especially regarding the relationship with children. There are indications of generational patterns of conflict, with one respondent reflecting on the impact of their parents' fighting during childhood and how it contributes to trust issues in their current relationships. The dynamics illustrated suggest a cycle of emotional abuse, particularly with references to demeaning comparisons among siblings and aggressive behaviour from parents, highlighting a pervasive culture of conflict that impacts self-esteem and relational stability. Overall, these insights underscore the complexities of familial relationships and the resulting emotional ramifications, leaving individuals feeling trapped and seeking guidance on navigational strategies for resolution.
\\ \hline

Anxieties and Self-Perception Struggles &
This cluster of responses reveals a deep-rooted struggle with self-acceptance and an overwhelming anxieties, particularly in social situations. The individuals express feelings of anxiety and self-loathing, often characterized by destructive behaviours like self-harm and emotional withdrawal. The fear of being judged by others significantly impacts their social interactions, leading to avoidance of crowded settings, including peer gatherings like school dances. The pervasive sense of loneliness and the belief that they are unworthy of love manifests in thoughts of being clingy and the difficulty in processing emotions beyond anxiety. A notable trend is the desire for self-improvement and confidence, coupled with a recognition of the need to confront and acknowledge their negative emotions rather than suppressing them. Overall, these responses highlight a cycle of anxiety, negative self-perception, and relational fears that contribute to a profound sense of isolation.
\\ \hline

Struggles with Trust and Toxic Relationships &
This cluster reveals deep emotional turmoil surrounding issues of trust, self-worth, and dependence in romantic relationships. Respondents express feelings of betrayal and heartbreak upon discovering their partner's inappropriate actions, leading to a profound sense of inadequacy and competition against others for their partner's affection. Many describe being caught in toxic dynamics characterized by emotional or verbal abuse, where the partner's manipulative behavior fosters feelings of helplessness and fear of being unable to find happiness without them. Concerns about the partner's dependency, such as being homeless or struggling with substance issues, further complicate their ability to leave an unhealthy relationship. The overarching theme is a struggle to navigate personal feelings of love and self-esteem amidst betrayal and abuse, alongside an urgent desire for clarity in deciding the future of these relationships.
\\ \hline

Emotional Turmoil in Romantic Relationships &
The responses in this cluster reveal a profound sense of emotional upheaval stemming from complex romantic entanglements. Key themes include the rapid escalation of intimacy followed by abrupt withdrawal, highlighting the fragility of affection in modern relationships. Individuals express deep love and connection, but also frustration and sorrow when faced with unexpected changes in their partners' commitment levels or emotional availability. A notable concern is the struggle between personal desires and external familial obligations, as evidenced by one partner's contemplation of rekindling a relationship for the sake of a child, despite lacking attraction. Patterns of betrayal, such as infidelity and lack of support during critical life events, further complicate these dynamics. There is a pervasive feeling of powerlessness and confusion, where individuals vacillate between the hope of rekindled love and the recognition of toxic patterns, leaving them uncertain about how to navigate their relationships moving forward.
\\ \hline

Struggles with Mental Health and Isolation &
The responses in this cluster reflect a shared narrative of persistent struggles with mental health, particularly depression, anxiety, and the long-lasting impacts of trauma. Many individuals express a sense of frustration regarding their inability to overcome anxiety and social isolation, despite some improvement in their depressive symptoms over time. There is a significant emphasis on feelings of hopelessness and the burden of hiding their struggles from family and friends, leading to feelings of exhaustion and loneliness. Concerns about stigma and misunderstandings from loved ones, particularly around the legitimacy of their mental health issues, highlight barriers to seeking help. Some individuals question whether they can manage their mental health challenges independently, while others express a desire for professional guidance, indicating a recognition of their need for support yet facing barriers such as affordability. Overall, the cluster underscores a cycle of mental health struggles, social disconnection, and the quest for acknowledgment and effective support.
\\ \hline

Sexual Confusion and Intimacy Challenges &
The responses in this cluster reflect a deep-seated struggle with sexual identity, performance, and intimacy, leading to considerable emotional distress. A recurring theme is the conflict between desire and aversion, evidenced by individuals grappling with unexpected sexual interests that disrupt their established sense of self, such as a man questioning his attractions after decades of heterosexuality or another who feels both excited and disgusted by the prospect of sharing his wife. There is also a significant focus on the impact of performance anxiety, with one participant noting that erectile dysfunction is eroding self-esteem and straining marital relations, reflecting how sexual health issues can influence emotional well-being and relational dynamics. Additionally, concerns about trust and safety surface in the context of BDSM, revealing a desire for exploration amidst feelings of anxiety and fear. Overall, this cluster highlights the complex interplay between sexual desires, emotional health, and relationship stability, emphasizing the need for open communication and therapeutic support.
\\ \hline

Trust and Infidelity Issues in Marriage &
This cluster of responses reveals a pervasive theme of trust issues and the impact of infidelity on marital relationships. Several individuals express feelings of betrayal, with partners either having cheated or harboring ongoing suspicions about fidelity. Emotional scars from past events, particularly surrounding the loss of a child, significantly influence one respondent's marital dynamics, leading to recurrent doubts about love and relationship stability during emotionally charged periods. The cyclical nature of these issues is evident, as couples navigate periods of reconciliation interspersed with emotional upheaval, often linked to unresolved grief or resentment. The responses also highlight personal struggles, such as health issues and mental wellbeing, that complicate recovery from infidelity. Ultimately, the common thread across the narratives is the challenge of rebuilding trust and the ongoing emotional toll inflicted by past betrayals, coupled with an underlying desire for healing and commitment despite the turmoil.
\\ \hline

\end{tabular}}
\end{table*}

\begin{table*}[ht]
\centering
\captionsetup{}
\caption{LLM Generated Cluster Labels and Descriptions for Human disclosures to Robot}
\label{tab:h2r_clusters}
\scalebox{0.85}{
\begin{tabular}{|p{1.5cm}|p{15cm}|}
\hline

\textbf{Cluster Label} & \textbf{Cluster Description} \\ \hline
Continuous Personal Development and Self-Reflection &
This cluster reveals a strong focus on personal development and self-reflection across various aspects of life, particularly in academic and professional contexts. Several responses highlight a deep enjoyment of the learning and growth process, wherein individuals derive joy and purpose from pushing themselves beyond their current limits, regardless of external outcomes like grades or results. There is a notable trend of seeking inspiration from others, whether through social skills, leadership qualities, or technical knowledge like programming. The experiences described suggest a blended emotional landscape where success in personal achievements fosters a sense of accomplishment, while challenges, such as frustrations with projects or mental health impacts, underscore the interconnectedness of emotions and productivity. Some individuals express a clear recognition of the need for better self-care and planning, especially as they transition into new phases in their academic or professional lives. The entrepreneurial experiences outlined further illustrate the value of resilience, networking, and skill acquisition in navigating the complexities of building a business. Overall, the responses indicate a proactive mindset towards self-improvement and adapting strategies in response to both internal and external challenges.
\\ \hline

Building Connections and Memorable Experiences &
The responses in this cluster emphasize the significance of social interactions and the creation of memorable experiences during the respondent’s time in Cambridge. Several themes emerge, including the joy derived from spending quality time with friends and family, as well as the appreciation for new relationships formed through shared activities. There is a noticeable trend of gratitude expressed towards new friendships, suggesting that these connections have positively impacted the individual's experience during what appears to be a transitional and potentially challenging period, such as a summer program. Activities like visits to local attractions, outings to restaurants, and participation in social events, such as cocktail-making classes, illustrate a proactive approach to building a support network. The references to enjoyable outings amidst familial and personal challenges further highlight the therapeutic power of shared moments and the importance of community and connection in enhancing personal well-being. Overall, the responses reflect a meaningful engagement with both the social landscape and the cultural offerings of Cambridge, fostering a sense of belonging and happiness.
\\ \hline

Academic Ambition and Future Aspirations &
Responses in this cluster reflect a strong sense of achievement, motivation, and future-focused ambition within the context of academic pursuits. The individual expresses pride in their accomplishments as a visiting researcher at a prestigious institution, demonstrating satisfaction with the opportunities afforded to them. There is a notable inclination towards setting and chasing new goals as soon as one is attained, suggesting a possibly insatiable drive for success. Moreover, there is a keen interest in the academic process itself, with enjoyment mentioned regarding the learning and writing involved in historical research. However, a tension arises from working on projects that may not fully align with personal research interests, leading to reflections on the long-term benefits of these experiences. Despite the struggles faced during their master’s program, the individual appears resolute in navigating these challenges as a means to define their niche. Ultimately, they express aspirations for an academic career that not only provides financial stability but also allows for personal interests and achievements to flourish, indicating a holistic view of professional fulfillment.
\\ \hline

Navigating Interpersonal Connections and Emotional Management &
The responses in this cluster reveal a significant focus on the complexities of interpersonal relationships and the emotional impact of disconnections. A common theme is the struggle between maintaining social ties and recognizing when those connections no longer align with personal values or interests, prompting a desire for distance. Participants express feelings of stress and overwhelm, particularly in managing multiple commitments like research projects and personal life, indicating a strain on their emotional well-being. They highlight a tendency to avoid confrontation, reflecting on past instances where unvoiced grievances led to feelings of sadness. This avoidance is coupled with a growing awareness of the importance of mindfulness and positivity as coping strategies, with some respondents mentioning the shift towards focusing on the positives in life to alleviate feelings of loneliness or stress. Overall, there is an acknowledgment of the impact of daily life experiences on emotional health, the continuous evolution of personal feelings, and the necessity of proactive emotional management and effective communication.
\\ \hline

Passion for Learning and Creativity &
The responses in this cluster reveal a strong passion for learning, exploration, and creativity, particularly through activities like reading and writing. The individual expresses a deep connection to personal pursuits that promote curiosity, indicating that they find fulfillment in engaging with subjects that intrigue them, whether it's through academic research or creative expressions such as storytelling. There is a notable contrast between fulfilling obligatory tasks that feel unworthy of their time versus immersive activities that resonate with their interests, highlighting a sense of frustration with tasks perceived as irrelevant. The individual draws satisfaction from intellectual challenges, such as understanding complex concepts in fields like artificial intelligence and neuroscience, suggesting they value the process of discovery and personal growth. Furthermore, moments of intense focus, like studying in the library, illustrate the joy derived from being absorbed in meaningful tasks, underscoring a desire to find significance and enjoyment in their learning experiences. Overall, the responses reflect a keen desire for engagement in pursuits that foster personal and professional development while grappling with societal expectations and perceived productivity.
\\ \hline

Friendships: Connection and Loneliness &
The responses in this cluster depict a nuanced exploration of friendships characterized by both meaningful connections and feelings of loneliness. Many participants express gratitude and happiness for deep conversations and shared experiences with friends, highlighting moments that reinforce their sense of belonging. There is a strong emphasis on the importance of communication and reconnection, as illustrated by anecdotes about catching up after periods of separation. However, contrasting these positive interactions are sentiments of alienation, particularly regarding relationships with housemates. Some individuals describe feelings of being misunderstood or not treated kindly, leading to loneliness in an environment where they expected camaraderie. This duality suggests a struggle between the desire for connection and the pain of feeling isolated, revealing that the quality of friendships greatly impacts emotional well-being. Overall, the responses emphasize the significance of nurturing and understanding relationships while acknowledging the complexities and challenges that can arise within them.
\\ \hline

\end{tabular}}
\end{table*}

\end{document}